\documentclass[pdflatex,sn-nature,Numbered,a4paper]{sn-jnl}

\usepackage{newtxtext,newtxmath}

\usepackage{graphicx}%
\usepackage{xcolor}%
\usepackage{manyfoot}%
\usepackage{booktabs}%
\graphicspath{{./figures/}}

\newcommand{\msun}{M$_\odot$}
\newcommand{\rsun}{R$_\odot$}
\newcommand{\kms}{km\,s$^{-1}$}
\newcommand{\obj}{WD\,0032$-$317}

\DeclareRobustCommand{\VAN}[3]{#2}
\let\VANthebibliography\thebibliography
\def\thebibliography{\DeclareRobustCommand{\VAN}[3]{##3}\VANthebibliography}

\begin{document}

\title[]{An irradiated-Jupiter analogue hotter than the Sun}








\author*[1]{\fnm{Na'ama} \sur{Hallakoun}}\email{naama.hallakoun@weizmann.ac.il}

\author[2]{\fnm{Dan} \sur{Maoz}}

\author[3]{\fnm{Alina~G.} \sur{Istrate}}

\author[4]{\fnm{Carles} \sur{Badenes}}

\author[5]{\fnm{Elm\'{e}} \sur{Breedt}}

\author[6]{\fnm{Boris~T.} \sur{G\"{a}nsicke}}

\author[7]{\fnm{Saurabh~W.} \sur{Jha}}

\author[8]{\fnm{Bruno} \sur{Leibundgut}}

\author[9]{\fnm{Filippo} \sur{Mannucci}}

\author[6]{\fnm{Thomas~R.} \sur{Marsh}\sfx{$^{\dag}$}}

\author[3,10,11]{\fnm{Gijs} \sur{Nelemans}}

\author[8]{\fnm{Ferdinando} \sur{Patat}}

\author[12,13]{\fnm{Alberto} \sur{Rebassa-Mansergas}}

\affil*[1]{\orgdiv{Department of particle physics and astrophysics}, \orgname{Weizmann Institute of Science}, \orgaddress{\city{Rehovot}, \postcode{7610001}, \country{Israel}}}

\affil[2]{\orgdiv{School of Physics and Astronomy}, \orgname{Tel-Aviv University}, \orgaddress{\city{Tel-Aviv}, \postcode{6997801}, \country{Israel}}}

\affil[3]{\orgdiv{Department of Astrophysics/IMAPP}, \orgname{Radboud University Nijmegen}, \orgaddress{\street{PO Box 9010}, \city{Nijmegen}, \postcode{6500 GL}, \country{the Netherlands}}}

\affil[4]{\orgdiv{Department of Physics and Astronomy and Pittsburgh Particle Physics, Astrophysics and Cosmology Center (PITT PACC)}, \orgname{University of Pittsburgh}, \orgaddress{\street{3941 O'Hara Street}, \city{Pittsburgh}, \postcode{15260}, \state{PA}, \country{USA}}}

\affil[5]{\orgdiv{Institute of Astronomy}, \orgname{University of Cambridge}, \orgaddress{\street{Madingley Road}, \city{Cambridge}, \postcode{CB3 0HA}, \country{UK}}}

\affil[6]{\orgdiv{Department of Physics}, \orgname{University of Warwick}, \orgaddress{\city{Coventry}, \postcode{CV4 7AL}, \country{UK}}}

\affil[7]{\orgdiv{Department of Physics and Astronomy}, \orgname{Rutgers, The State University of New Jersey}, \orgaddress{\street{136 Frelinghuysen Road}, \city{Piscataway}, \postcode{08854}, \state{NJ}, \country{USA}}}

\affil[8]{\orgname{European Southern Observatory}, \orgaddress{\street{Karl-Schwarzschild-Stra{\ss}e 2}, \city{Garching}, \postcode{D-85748}, \country{Germany}}}

\affil[9]{\orgname{INAF -- Osservatorio Astrofisico di Arcetri}, \orgaddress{\street{Largo E. Fermi 5}, \city{Firenze}, \postcode{I-50125}, \country{Italy}}}

\affil[10]{\orgdiv{Institute for Astronomy}, \orgname{KU Leuven}, \orgaddress{\city{Celestijnenlaan 200D, B-3001 Leuven}, \country{Belgium}}}

\affil[11]{\orgdiv{SRON}, \orgname{Netherlands Institute for Space Research}, \orgaddress{\street{Niels Bohrweg 4}, \city{Leiden}, \postcode{NL-2333 CA}, \country{the Netherlands}}}

\affil[12]{\orgdiv{Departament de F\'{i}sica}, \orgname{Universitat Polit\`{e}cnica de Catalunya}, \orgaddress{\street{c/Esteve Terrades 5}, \city{Castelldefels}, \postcode{E-08860}, \country{Spain}}}

\affil[13]{\orgdiv{Institut d'Estudis Espacials de Catalunya}, \orgname{Ed. Nexus-201}, \orgaddress{\street{c/Gran Capit\`{a} 2-4}, \city{Barcelona}, \postcode{E-08034}, \country{Spain}}}

\affil[\dag]{Deceased}


\abstract{Planets orbiting close to hot stars experience intense extreme-ultraviolet radiation, potentially leading to atmosphere evaporation and to thermal dissociation of molecules. However, this extreme regime remains mainly unexplored due to observational challenges.
Only a single known ultra-hot giant planet, KELT-9b, receives enough ultraviolet radiation for molecular dissociation, with a day-side temperature of $\approx4,600$\,K.
An alternative approach uses irradiated brown dwarfs as hot-Jupiter analogues. With atmospheres and radii similar to those of giant planets, brown dwarfs orbiting close to hot Earth-sized white-dwarf stars can be directly detected above the glare of the star.
Here we report observations revealing an extremely irradiated low-mass companion to the hot white dwarf \obj. Our analysis indicates a day-side temperature of $\approx 8,000$\,K, and a day-to-night temperature difference of $\approx 6,000$\,K. The amount of extreme-ultraviolet radiation (with wavelengths $100-912$\,\AA) received by \obj B is equivalent to that received by planets orbiting close to stars as hot as a late B-type stars, and about $5,600$ times higher than that of KELT-9b.
With a mass of $\approx 75-88$ Jupiter masses, this near-hydrogen-burning-limit object is potentially one of the most massive brown dwarfs known.}

\maketitle

\section{Introduction}

When a planet orbits very close to a star, the strong tidal forces it experiences tend to synchronise its orbital and rotational periods, permanently locking one side of the planet facing the star (`tidal locking'). The planet's `day-side' hemisphere is then continuously exposed to direct radiation. Depending on the heat redistribution on the planet surface, this can lead to extreme temperature differences between the day and night sides of the planet, and to thermal dissociation  of the molecules on the planet's day side \cite{Kitzmann_2018, Lothringer_2018}. Out of the few dozen ultra-hot giant planets discovered so far \cite{Akeson_2013}, only KELT-9b receives ultraviolet radiation high enough in amount for molecular dissociation, with a day-side temperature of $\approx4,600$\,K \cite{Gaudi_2017}.

Our knowledge of planetary systems around hot massive stars is extremely limited. These stars have few spectral lines, which are significantly broadened by their rapid rotation and by stellar activity \cite{Wright_2005}, making high-precision radial-velocity measurements challenging. Such measurements are crucial for planet detection and confirmation, and hence known planets are scarce around stars more massive than $\sim 1.5$\,\msun. The difficulty in detecting ultra-hot Jupiters and directly examining their atmospheres limits our ability to test theoretical atmospherical models.

An alternative approach uses irradiated brown dwarfs as hot-Jupiter analogues \cite{Lee_2020, Lothringer_2020, Tan_2020}. Despite being more massive than giant planets, brown dwarfs have comparable sizes. Binary systems consisting of a brown dwarf and a white dwarf \cite[e.g.][]{vanRoestel_2021} are of particular interest, as intense irradiation by a hot white dwarf is possible due to the small radius of the white dwarf which permits very close companion orbits without contact. At the same time, the same small sizes of white dwarfs (with radii an order of magnitude smaller than those of brown dwarfs) makes them many orders of magnitude less luminous than massive stars, revealing the companion above the glare of the star. Since the host white dwarf is much hotter than the brown dwarf, it also dominates the light at different ranges of the electromagnetic spectrum---white dwarfs emit mostly in the ultraviolet and optical regions, while brown dwarfs emit mostly in the infrared.

\obj\ is a hot ($\approx 37,000$\,K) low-mass ($\approx 0.4$\,\msun) white dwarf. Its high effective temperature indicates that only $\sim 1$~million years (Myr) have passed since its progenitor star became a white dwarf. High-resolution spectra of the object were obtained in the early 2000's during the Type-Ia Supernova Progenitor surveY (SPY) \cite{Napiwotzki_2020}, that was carried using the Ultra-Violet-Visual Echelle Spectrograph (UVES) \cite{Dekker_2000} of the European Southern Observatory (ESO) Very Large Telescope (VLT) at Paranal, Chile. These data showed a significant radial-velocity shift of its hydrogen H$\alpha$ absorption line, caused by the reflex motion induced by the presence of a close companion, flagging \obj\ as a potential double white dwarf system in the candidate list of Maoz and Hallakoun \cite{Maoz_2017}. A weak near-infrared excess in the archival spectral energy distribution of \obj\ noted in \cite{Maoz_2017}, hinted that the companion could actually be a brown dwarf rather than another white dwarf.

\section{Results}

New follow-up data that we have obtained with UVES, in settings similar to the original SPY spectra, reveal the presence of a highly-irradiated low-mass companion, evident by the presence of Balmer emission lines in anti-phase with the primary white dwarf absorption lines (Fig.~\ref{fig:Halpha} and Extended Data Figs.~\ref{fig:Halpha1D}--\ref{fig:FullSpectra}). The companion's emission in this tidally-locked system is only detected when its heated day side is facing us, while the radiation coming from the cooler night-side hemisphere remains hidden in the glare of the white dwarf in the observed wavelength range. The original SPY spectra were fortuitously obtained when the companion's night side was visible, hiding the day-side emission. We have extracted and fitted the radial-velocity curves of the white dwarf and companion, and found an orbital period of about 2.3~hours (see Table~\ref{tab:param} and Extended Data Figs.~\ref{fig:LS} and \ref{fig:MCMC}).
We only detect hydrogen emission lines from the companion, similarly to other systems with highly-irradiated companions \cite{Farihi_2017, Parsons_2017, Schaffenroth_2021}, although we note that emission lines from metals have been detected in other similar systems \cite{Parsons_2010, Longstaff_2017, Casewell_2018}.

\subsection{Determining the white dwarf mass}
In order to convert the radial-velocity fit parameters into the physical properties of the system, we need to assume a mass for the white dwarf. The effective temperature and the surface gravity of the white dwarf (Table~\ref{tab:param}) have been previously estimated based on an atmospheric fit to the original SPY UVES observations in 2000 \cite{Koester_2009}. These parameters can be converted into a mass, a radius, and a cooling age using theoretical evolutionary tracks, by assuming a specific white-dwarf core composition.
While `normal' white dwarfs have cores composed of carbon and oxygen, white dwarfs with masses below $\sim 0.45$\,\msun\ are considered low-mass white dwarfs, and could not have formed via single-star evolution, since their progenitor main-sequence lifetime is longer than the age of the Galaxy. Such white dwarfs are generally thought to have helium cores, a result of their nucleosynthetic evolution having been truncated by binary interactions (e.g. \cite{Althaus_2013}). Alternatively, if the white dwarf mass is not extremely low ($\sim 0.3$\,\msun), intermediate-mass progenitors ($\gtrapprox~2.1$\,\msun) in binary systems (or undergoing extreme mass loss through stellar winds) can leave behind a hybrid-core white dwarf, i.e. a carbon-oxygen core surrounded by a thick helium layer (e.g. \cite{Iben_1985, Zenati_2019, Romero_2021}).
Since the mass of \obj\ is in the low-mass range ($\approx 0.4$\,\msun), we have considered the implications of assuming helium- (He) and hybrid-core white dwarfs in our analysis.

\subsection{Fitting the spectral energy distribution of the system}
To look for photometric variability, we obtained photometric data in multiple wavelength bands using the 1-m Las Cumbres Observatory Global Telescope (LCOGT) network \cite{Brown_2013}. In addition, we retrieved archival light curves from NASA's Transiting Exoplanet Survey Satellite (TESS) \cite{Ricker_2014} and Wide-field Infrared Survey Explorer (WISE) \cite{Wright_2010}. The light curves show a clear sinusoidal modulation resulting from the changing phases, from the observer's viewpoint, of the irradiated hemisphere of the companion. The photometric period is consistent with the one obtained from the radial-velocity curves, with no detected eclipses (Extended Data Figs.~\ref{fig:LC} and \ref{fig:LC_LS}).

\begin{figure}[h!]
    \centering
    \includegraphics[width=0.6\textwidth]{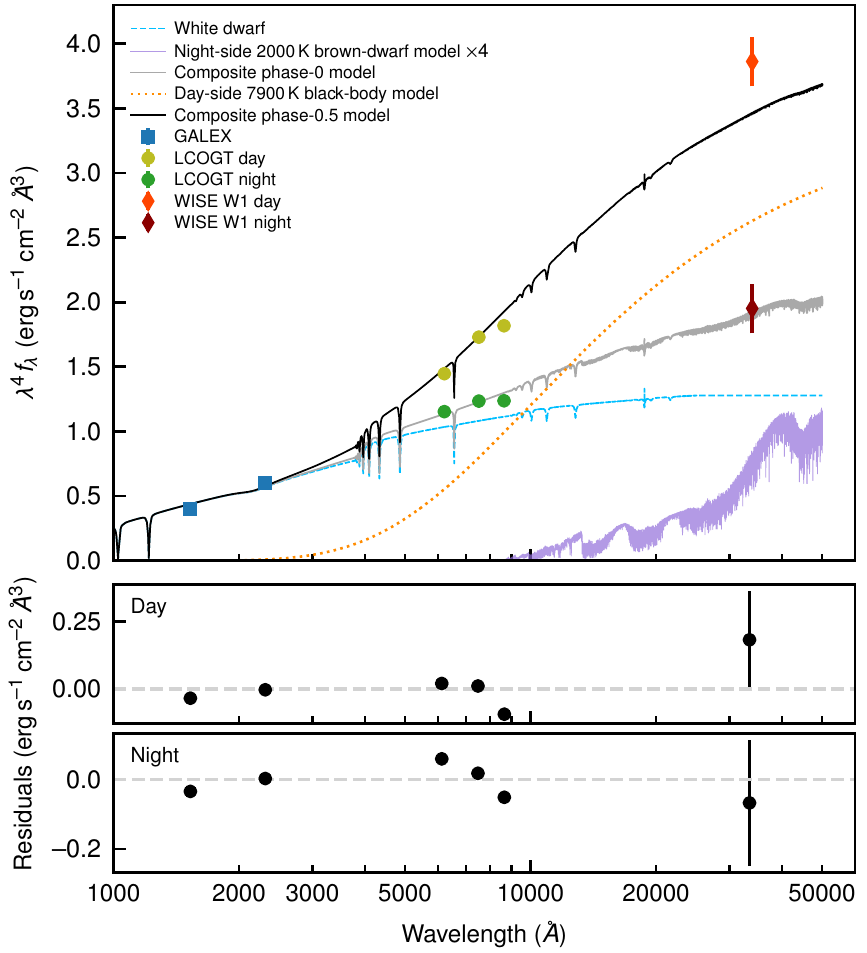}
    \caption{\textbf{Observed spectral energy distribution for \obj}, compared to the best-fitting composite theoretical model spectra of a white dwarf and a black body/brown dwarf. The archival GALEX ultraviolet photometry, where the contribution from the companion is negligible, appears as blue square-shaped error bars.
    Minimal/maximal photometric values in different bands, extracted from the light curves, appear as green-shades circle-shaped error bars for LCOGT's $r'$, $i'$, and $z$ bands, and as red-shades diamond-shaped error bars for the WISE W1 band.
    A theoretical model spectrum of a hydrogen-dominated white dwarf with an effective temperature of $37,000$\,K and a surface gravity $\log g=7.2$ \cite{Levenhagen_2017} is shown in dashed light blue. The best-fitting brown-dwarf (\cite{Allard_2011, Allard_2012}; for the night side, with $[\textrm{M/H}]=-0.5$ and $\log g=5.5$) and black-body (for the day side) models are plotted in solid purple and dotted orange, respectively. The theoretical spectra were scaled using the system's distance measured by the Gaia mission, and the estimated component radii (assuming a He-core white dwarf, see Extended Data Fig.~\ref{fig:SED} for the hybrid model). The brown-dwarf model is shown multiplied by a factor of 4, to fit the displayed range.
    The composite model of the system at orbital phase 0 (0.5) is plotted in solid dark grey (black).
    The units shown on the $y$ axis are the flux per wavelength, $\lambda$, multiplied by $\lambda^4$, for visual clarity.
    The bottom panels show the residuals of the day-side (\textit{middle}) and the night-side (\textit{bottom}) fits. The error bars in the residual plots show the standard deviation and take into account both the photometric and the model uncertainties.}
    \label{fig:SED_He}
\end{figure}

We have estimated the companion's radius, as well as its night- and day-side effective temperatures, by fitting the spectral energy distribution of the system with a combination of a white-dwarf model spectrum and a brown-dwarf model spectrum for the cooler night side, and with a black-body spectrum for the day side (Fig.~\ref{fig:SED_He} and Extended Data Figs.~\ref{fig:SED}, \ref{fig:MCMC_SED_He}, and \ref{fig:MCMC_SED_Hybrid}). We note that the actual day-side spectrum of \obj\ is not expected to exactly follow that of a black body, since different wavelength ranges probe different optical depths with different pressures \cite{Zhou_2022}. To account for the system's orbital inclination we have included an additional fitting parameter indicating the fraction of night/day contamination.
Depending on the white-dwarf core model used, the companion's heated day-side temperature ranges between $\approx 7,250$ and $9,800$\,K---as hot as an A-type star---with a night-side temperature of $\approx1,300-3,000$\,K, or a temperature difference of $\approx 6,000$\,K---about four time as large as that of KELT-9b \cite{Wong_2020}. The night-side temperature range covers T through M dwarfs.
The `equilibrium' black-body temperature of the irradiated companion (neglecting its intrinsic luminosity and albedo, and assuming it is in thermal equilibrium with the external irradiation) is about $5,100$\,K, hotter than any known giant planet (Fig.~\ref{fig:WDBD}), and $\approx 1,000$\,K hotter than KELT-9b \cite{Gaudi_2017}, resulting in $\approx5,600$ times higher extreme-ultraviolet flux.
We note that the irradiated companion of the hot white dwarf NN\,Serpentis has an even higher equilibrium temperature of $\approx 6000$\,K \cite{Parsons_2010} (but only about three times the amount of extreme-ultraviolet radiation received by \obj B). However, with a mass of $0.111\pm0.004$\,\msun\ the companion of NN\,Serpentis is a bona fide main-sequence star rather than a brown dwarf or a near hydrogen-burning limit object (see Fig.~\ref{fig:MRR}).

\subsection{Near-infrared spectroscopy}
We obtained a pair of low-resolution near-infrared spectra using the Gemini South's FLAMINGOS-2 spectrograph \cite{Eikenberry_2004}, taken near orbital phases 0 and 0.35 (Extended Data Fig.~\ref{fig:F2}). As expected \cite{Zhou_2022}, the slope of the spectra at this wavelength range is dominated by the irradiated hemisphere's black-body tail at all orbital phases (because of the relatively low inclination of the system). However, due to the low signal-to-noise ratio and possible telluric contamination, we could not confidently identify any finer features, that are expected at the few-percent level in this wavelength range. At orbital phase 0.35, when a larger fraction of the irradiated hemisphere is visible, a possible weak Brackett $10 \rightarrow 4$ hydrogen line emission is detected. Future infrared spectroscopic observations with high single-to-noise ratio (e.g. with the James Webb Space Telescope), taken at different orbital phases, should be able to resolve these features.

\section{Discussion}
\begin{figure*}[h!]
    \centering
    \includegraphics[width=\textwidth]{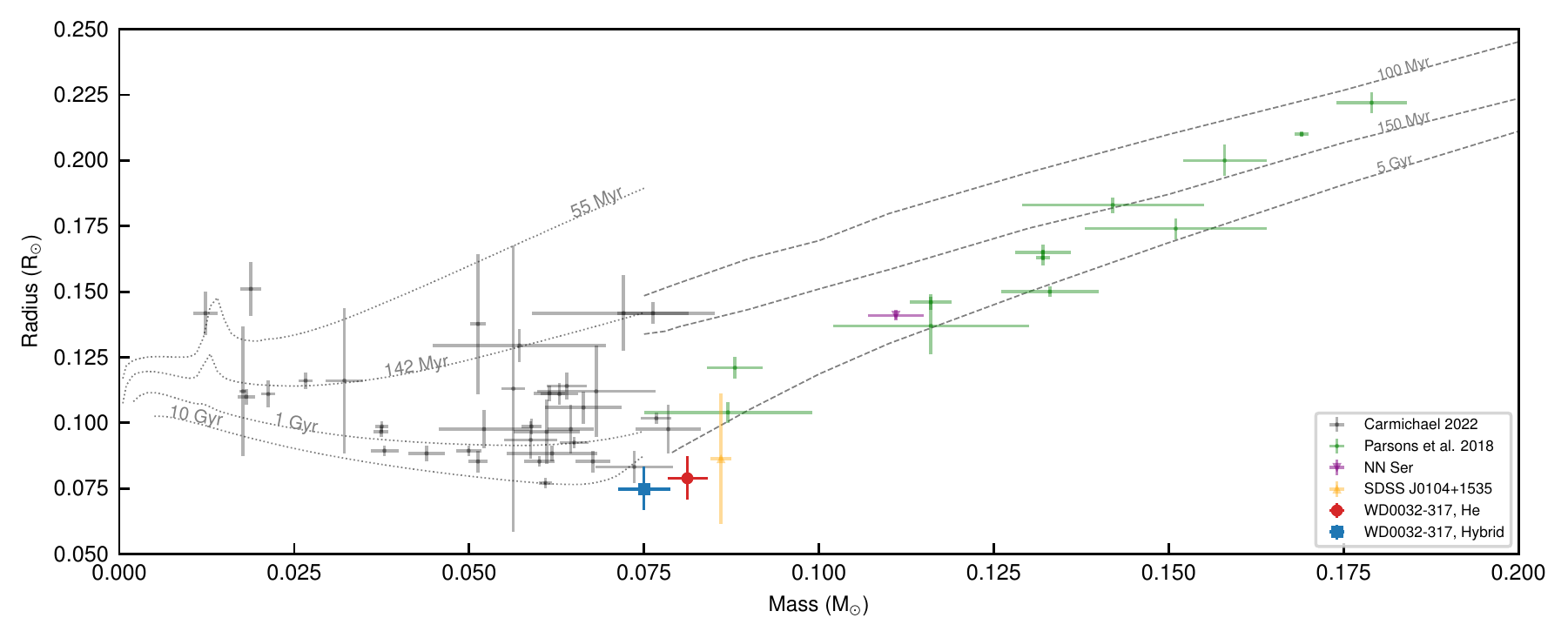}
    \caption{\textbf{Brown-dwarf and low-mass stars mass-radius relation.} Known transiting brown-dwarfs and very low-mass stars orbiting main-sequence stars appear as grey error bars \cite{Carmichael_2022}, while known eclipsing M-dwarfs orbiting white dwarfs appear as green error bars \cite{Parsons_2018}. The dotted grey lines are theoretical ATMO2020 isochrones \cite{Phillips_2020} of different ages and solar metallicity. The dashed grey lines are theoretical BT-DUSTY isochrones \cite{Allard_2011, Allard_2012} of different ages, with $[\textrm{M/H}]=-0.5$. The position of \obj\ is marked by a red circle (assuming a He-core white dwarf), and a blue square (assuming a hybrid-core white dwarf). The low-metallicity halo object, SDSS\,J0104+1535, which is the most massive known brown dwarf \cite{Zhang_2017}, is shown in orange triangle, and the ultra-hot M-dwarf companion of NN\,Ser \cite{Parsons_2010} is shown in purple down-pointing triangle, for reference. The error bars show the standard deviation.}
    \label{fig:MRR}
\end{figure*}

The main source of uncertainty in determining the properties of the system remains the white-dwarf core composition, with the companion mass ranging from $\approx 0.075$\,\msun\ for a hybrid-core white dwarf, and $\approx 0.081$\,\msun\ for a He-core white dwarf, both near the hydrogen-burning limit. Although theoretical evolutionary models place this limit somewhere between $0.070-0.077$\,\msun\ for solar metallicity, observations suggest a higher limit \cite{Dieterich_2014, Chabrier_2022}. Since the precise hydrogen-burning limit depends on the metallicity \cite{Chabrier_1997}, rotation \cite{Chowdhury_2022}, and formation history of the brown dwarf \cite{Forbes_2019}, the companion could still be a very massive brown dwarf. Inconsistencies between the predicted theoretical mass and the much-higher measured dynamical mass of some T dwarfs have also been reported \cite{Brandt_2020}. The three-dimensional velocity of the system, $\approx 50$\,\kms, indicates a somewhat older age than that of the Galactic thin disc, which might point to a relatively lower metallicity. When placed on a mass-radius relation diagram (Fig.~\ref{fig:MRR}) it is clear that \obj\,B is a borderline object, with a smaller radius than expected for a non-degenerate hydrogen-burning star. Nevertheless, since at this mass range near the hydrogen-burning limit its intrinsic luminosity is negligible compared to the external radiation it experiences, the difference between a brown dwarf and a very low-mass star is merely semantic for the purpose of studying highly irradiated sub-stellar objects and planets.

In order to form the low-mass white dwarf, the companion must have contributed to the unbinding of the red giant's envelope. With a mass well above the critical limit of $\approx 0.01-0.03$\,\msun\ in the case of a He-core white dwarf, the companion was massive enough to have survived the process without getting evaporated \cite{Nelemans_1998}. The small radius of the companion, indicating an age of at least a few billion years (Gyr; Fig.~\ref{fig:MRR}), stands in contrast with the white-dwarf $\sim 1$\,Myr cooling age---the time that has passed since it lost its envelope. This suggests that the companion was not significantly heated during the common-envelope phase, indicating a rather efficient envelope ejection. Assuming the full energy required to unbind the envelope came from orbital sources, the progenitor of a He-core white dwarf could have been quite a low-mass star of $\sim 1.3$\,\msun\ \cite{Nelemans_1998}. Hybrid-core white dwarfs, on the other hand, are the descendants of more massive and compact giants, with much larger binding energies (e.g. \cite{Hu_2007}). This would require unbinding the envelope with a much higher efficiency in order for the companion to survive and get to the observed close orbit, and might argue against a hybrid nature of the white dwarf (see Methods).

\obj\ offers a rare glimpse into the early days of a post-common-envelope binary, and to an unexplored parameter space of irradiated substellar and planetary objects. Unlike actual hot Jupiters or irradiated brown dwarfs with larger host stars (such as hot subdwarfs, e.g. \cite{Schaffenroth_2021}), for which spectroscopic observations are only possible during eclipses in eclipsing systems, the low-mass companion should be visible in the infrared wavelength range throughout the orbital cycle. Future high-resolution time-resolved spectroscopic observations of the system covering the near-infrared range would reveal in detail the gradual transition from the absorptive feature-rich night side to the emissive day side (e.g. \cite{Zhou_2022}; Extended Data Fig.~\ref{fig:F2}), directly probing the effects of the extreme temperature difference and heat transport efficiency between the hemispheres. The broad wavelength coverage, sensitive to different pressure levels in the atmosphere, would reveal the three-dimensional atmospheric structure, including temperature inversion effects \cite{Casewell_2015, Zhou_2022}. Since the system is tidally locked, the orbital period provides a direct measurement of the companion rotation period. This can help in understanding the role of rotation on the atmospheric structure and circulation in fast-rotating extremely-irradiated gas giants \cite{Tan_2020}.

\begin{table}[h!]
\begin{center}
\begin{minipage}{\textwidth}
\caption{Properties of the \obj\ system}\label{tab:param}
\begin{tabular*}{\textwidth}{@{}p{0.03\linewidth}p{0.5\linewidth}@{ }p{0.2\linewidth}@{ }p{0.2\linewidth}@{}}
\toprule
\multicolumn{4}{@{}l@{}}{\textbf{General system parameters}} \\
\midrule
\multicolumn{4}{@{}l@{}}{
\begin{tabular}{@{}p{0.1\linewidth}p{0.64\linewidth}@{ }p{0.2\linewidth}@{}}
RA & Right ascension (J2000)\footnotemark[1] & 00h34m49.8573s\\
Dec & Declination (J2000)\footnotemark[1] & $-31^{\circ}29'52.6858''$\\
$\varpi$ & Parallax\footnotemark[1] (mas) & $2.320\pm0.053$ \\
$d$ & Distance\footnotemark[1] (pc) & $431.1\pm9.8$ \\
$E \left( B-V \right)$ & Extinction\footnotemark[2] (mag) & $0.0176\pm0.0007$ \\
\end{tabular}
}\\
\midrule
\multicolumn{4}{@{}l@{}}{\textbf{White dwarf parameters\footnotemark[3]}} \\
\midrule
$T_1$ & \multicolumn{2}{l}{Effective temperature (K)} & $36,965\pm100$\\
$\log g_1$ & \multicolumn{2}{l}{Surface gravity (cm\,s$^{-2}$)} & $7.192\pm0.014$ \\
\midrule
\multicolumn{4}{@{}l@{}}{\textbf{Model-independent orbital parameters\footnotemark[4]}} \\
\midrule
$P$ & \multicolumn{2}{l}{Orbital period (s)} & $8340.9090 \pm 0.0013$ \\
$K_1$ & \multicolumn{2}{l}{Primary radial velocity semi-amplitude (\kms)} & $53.4 \pm 1.7$ \\
$K_\textrm{em}$ & \multicolumn{2}{l}{Secondary's emission radial velocity semi-amplitude (\kms)} & $257.1 \pm 1.1$ \\
$\gamma_1$ & \multicolumn{2}{l}{Primary mean velocity (\kms)} & $20.5 \pm 1.4$ \\
$\gamma_2$ & \multicolumn{2}{l}{Secondary mean velocity (\kms)} & $9.1 \pm 1.0$ \\
$\Delta \gamma$ & \multicolumn{2}{l}{Mean velocity difference (\kms)} & $11.4 \pm 1.7$ \\
$\phi_0$ & \multicolumn{2}{l}{Initial orbital phase} & $0.000^{+0.012}_{-0.011}$ \\
$T_0$ & Ephemeris (BJD (TDB); $E$ is the cycle number) & \multicolumn{2}{@{}r@{}}{$2451803.6673(11) + 0.096531354(15) E$}\\
\midrule
\multicolumn{4}{@{}l@{}}{\textbf{Model-dependent orbital parameters}} \\
\midrule
& & \multicolumn{2}{@{}c@{}}{White-dwarf core model} \\
\cmidrule{3-4}
 & & He\footnotemark[5] & Hybrid\footnotemark[6] \\
\midrule
$M_1$ & Primary mass (\msun) & $0.4187\pm0.0047$ & $0.386\pm0.014$ \\
$R_1$ & Primary radius (\rsun) & $0.02703\pm0.00024$ & $0.02616\pm0.00024$ \\
$t_1$ & Primary cooling age (Myr) & $0.91\pm0.30$ & $1.8\pm1.6$ \\
$M_2$ & Secondary mass (\msun) & $0.0812\pm0.0029$ & $0.0750\pm0.0037$ \\
$R_2$ & Secondary radius (\rsun) & $0.0789_{-0.0083}^{+0.0085}$ & $0.0747_{-0.0079}^{+0.0085}$ \\
$q$ & Mass ratio & $0.1939\pm0.0065$ & $0.1943\pm0.0065$ \\
$K_2$ & Secondary radial velocity semi-amplitude (\kms) & $275.6\pm2.4$ & $275.1\pm2.5$ \\
$a$ & Orbital separation (\rsun) & $0.7028\pm0.0026$ & $0.6841\pm0.0083$ \\
$i$ & Orbital inclination (deg) & $63.3\pm1.1$ & $66.4\pm2.0$ \\
$f_\text{cont}$ & Night/day contamination fraction & $0.182_{-0.034}^{+0.033}$ & $0.227_{-0.028}^{+0.028}$ \\
$T_\textrm{eq}$ & Secondary equilibrium temperature (K) & $5126\pm28$ & $5111\pm41$ \\
$T_\textrm{night}$ & Secondary night-side temperature (K) & $1970_{-670}^{+840}$ & $2035_{-716}^{+927}$ \\
$T_\textrm{day}$ & Secondary day-side temperature (K) & $7900_{-650}^{+780}$ & $8835_{-794}^{+955}$ \\
\botrule
\end{tabular*}
\footnotetext{Source:}
\footnotetext[1]{Gaia DR3}
\footnotetext[2]{\href{https://irsa.ipac.caltech.edu/applications/DUST/}{https://irsa.ipac.caltech.edu/applications/DUST/} \cite{Schlafly_2011}}
\footnotetext[3]{Atmospheric fit \cite{Koester_2009}}
\footnotetext[4]{Radial-velocity fit}
\footnotetext[5]{Helium-core white dwarf evolutionary tracks}
\footnotetext[6]{Hybrid-core white dwarf evolutionary tracks}
\footnotetext{Data are presented as median values $\pm$ standard deviation.}
\end{minipage}
\end{center}
\end{table}

\begin{figure*}[h!]
    \centering
    \includegraphics[width=\textwidth]{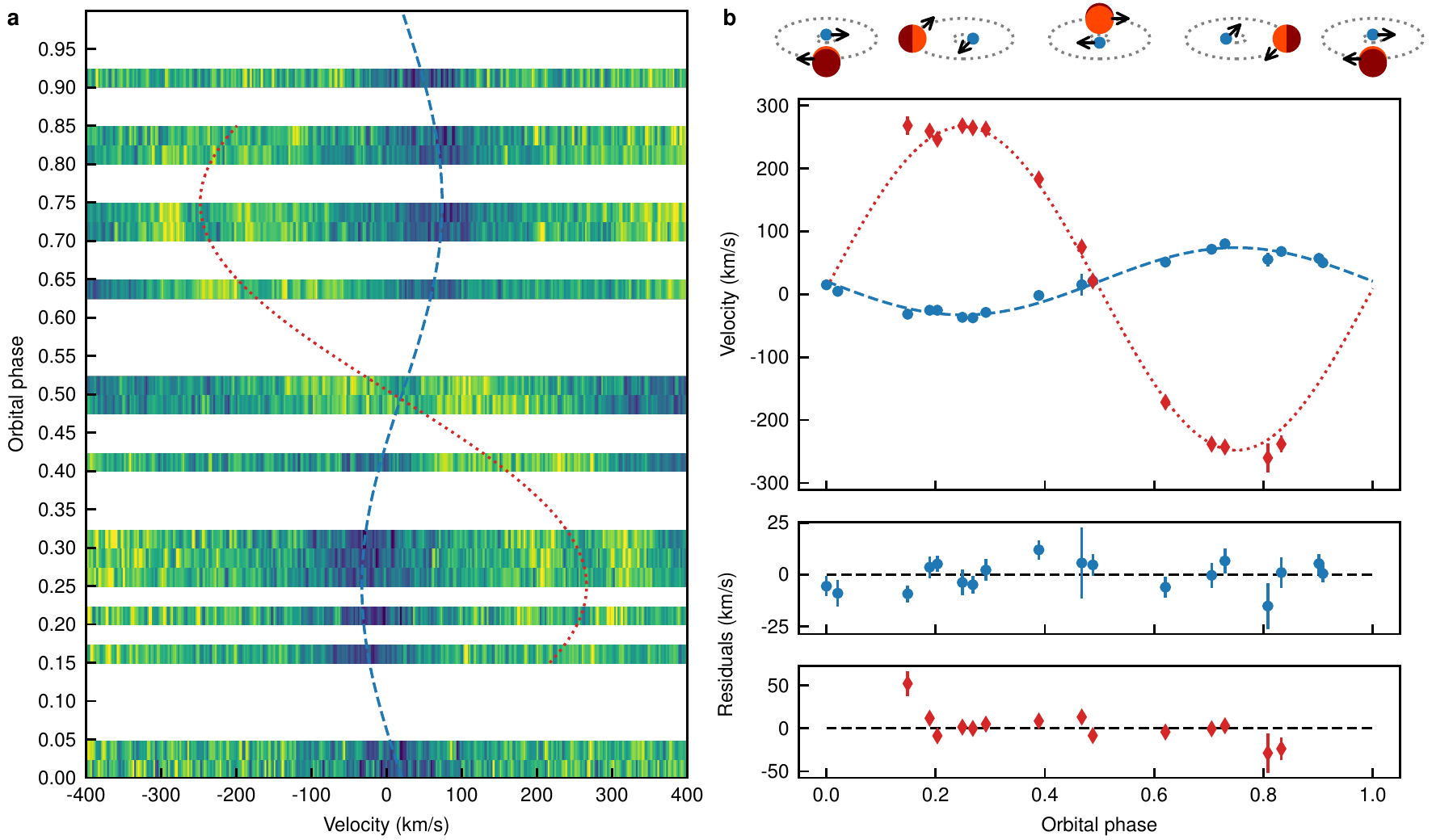}
    \caption{\textbf{Phased radial-velocity curves of \obj.}
    \textbf{a}, trailed UVES spectrum for the H$\alpha$ line of \obj\ (blue represents lower fluxes, and yellow represents higher fluxes), folded over the orbital period ($P=8340.9090$\,s). The primary absorption is clearly seen in blue. The emission from the companion (in yellow) appears in anti-phase with the primary, and is visible only from the irradiated day side, between orbital phases $\sim0.2-0.8$. Its `inverted' shape, evident especially near quadrature, is the result of non-local thermodynamic equilibrium (NLTE) effects \cite{Barman_2004}.
    \textbf{b}, radial velocity curves (\textit{top panel}) of the white dwarf (blue circles) and the irradiated companion (red diamonds), folded over the orbital period ($P=8340.9090$\,s). The primary's (secondary's) best-fit curve is marked by the blue dashed (red dotted) line on both panels. The bottom panels show the residuals of the white-dwarf component (\textit{middle}) and the irradiated companion (\textit{bottom}). The error bars show the standard deviation. The illustrations on \textit{top} demonstrate the system's configuration at each orbital phase.}
    \label{fig:Halpha}
\end{figure*}

\begin{figure}[h!]
    \centering
    \includegraphics[width=0.6\textwidth]{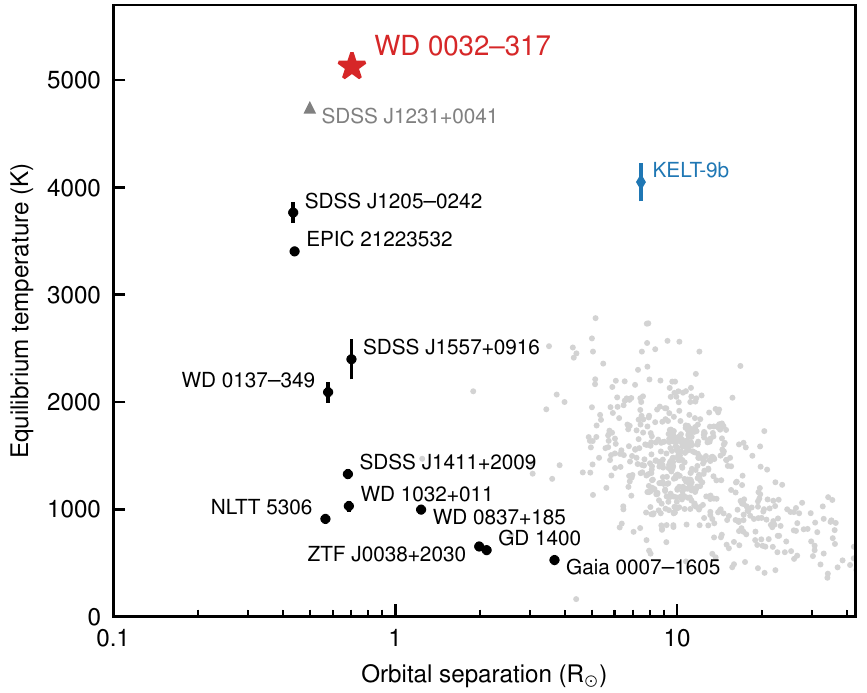}
    \caption{\textbf{The equilibrium temperature of \obj\ compared to other known systems.} Equilibrium temperature as a function of the orbital separation for the known (black circles) and candidate (grey triangle) white dwarf + brown dwarf systems (see Methods for references) and hot Jupiter planets (light-grey circles)\cite{Akeson_2013}. \obj\ is marked with a red star-shaped symbol. The ultra-hot Jupiter KELT-9b \cite{Gaudi_2017} appears as a blue diamond. The error bars show the standard deviation, and are plotted for all the objects (but are smaller than the marker size in some).}
    \label{fig:WDBD}
\end{figure}

\clearpage

\section*{Methods}\label{sec:methods}

\subsection*{Spectroscopic observations with UVES}
The two original 10-min exposures of \obj\ were obtained on 2000 September 16--17 with the UV-Visual Echelle Spectrograph (UVES) \cite{Dekker_2000} of the European Southern Observatory (ESO) Very Large Telescope (VLT) at Paranal, Chile, as a part of the SPY programme. The instrument was used in a dichroic mode, covering most of the range between $3200$\,\AA\ and $6650$\,\AA, with two $\sim 80$\,\AA\ gaps around $4580$\,\AA\ and $5640$\,\AA, and a spectral resolution of $R \approx 18,500$ ($0.36$\,\AA\ at H$\alpha$) \cite{Napiwotzki_2003}. The data featured Balmer absorption lines from the primary white dwarf only, with no contribution from a companion. In a previous work, we measured a radial-velocity shift of $38.1 \pm 3.8$\,\kms\ between the two epochs, and flagged it for future followup as a potential double white dwarf, noting that a weak infrared excess in the spectral energy distribution could indicate the presence of a brown-dwarf companion \cite{Maoz_2017}. We have acquired an additional 16 spectra with UVES using similar settings, from June to August 2019, and from September to December 2020. The pipeline-processed reduced data were obtained from the ESO Science Archive Facility. The new spectra revealed the presence of a highly-irradiated low-mass companion, evident in Balmer emission lines at anti-phase with the primary white dwarf (see Fig.~\ref{fig:Halpha}a and Extended Data Figs.~\ref{fig:Halpha1D} and \ref{fig:FullSpectra}).

\subsection*{Photometric observations with LCOGT}
To look for photometric variability, we obtained multi-band photometry on 2021 August 2--6 and 2022 October 14 using the 1-m Las Cumbres Observatory Global Telescope (LCOGT) network \cite{Brown_2013}. The observations were performed using the Sinistro cameras on the 1\,m-telescopes in Cerro Tololo (LSC), Chile, and in Sutherland (CPT), South Africa. Each observation sequence consisted of 50\,s exposures with an $\approx 1$\,min cadence, spanning $\sim 1-2$ orbital cycles, each sequence using a different filter (SDSS \textit{r'} and \textit{i'}, and Pan-STARRS \textit{z} and \textit{y}). The system was not detected in the \textit{y}-band images, and we hence omit the \textit{y}-band from the discussion.

The images were reduced by the BANZAI pipeline \cite{McCully_2018}, including bad-pixel masking, bias and dark subtraction, flat-field correction, source extraction, and astrometric calibration (using \url{http://astrometry.net/}). The source extraction was performed using \textsc{sep}, the \textsc{Python} and \textsc{C} library for Source Extraction and Photometry \cite{Bertin_1996, Barbary_2016, Barbary_2016b}. We then chose sources with signal-to-noise ratio between 100 and 1000 (to avoid faint and saturated stars, respectively), and flux standard deviation smaller than 30 times the mean flux error (to avoid light curves with long-term trends), as comparison stars. The raw light curves of the target and of a nearby reference star (RA=00h35m02.2571s, Dec=$-31^{\circ}31'19.028''$) were corrected for transparency variations by dividing them by the median flux of the comparison stars. The target light curve was then flux-calibrated using synthetic photometry extracted from the low-resolution BP/RP spectra of the target and reference stars in the third data release of Gaia \citep{Gaia_2016, Gaia_2022summary, Gaia_2022synthphot}, taking into account the colour difference between the two stars: We divided the corrected light curve of the target by that of the reference star, and multiplied it by the reference star flux in each band (calculated using the Gaia spectrum) and by a calibration factor that keeps the median count ratio equal to the Gaia synthetic photometry ratio in the band. The timestamps were shifted to mid-exposure and transformed to the barycentric frame using \textsc{astropy}.

\subsection*{Photometric observations with TESS and WISE}
The system was observed twice by the Transiting Exoplanet Survey Satellite (TESS) \cite{Ricker_2014}. Once in short-cadence mode (120\,s exposures) in 2018 from August 23rd to September 20th (Sector 02), and again in fast-cadence mode (20\,s exposures) in 2020 from August 26th to September 21st (Sector 29).
The pipeline-reduced light curves were obtained from the Mikulski Archive for Space Telescopes (MAST).

Additional 27 epochs of the system in the 3.4\,\textmu m $W1$ band were obtained by the Wide-field Infrared Survey Explorer (WISE) satellite \cite{Wright_2010} in 2010. The pipeline-reduced light curve was obtained from the Infrared Processing and Analysis Center (IPAC) Infrared Science Archive (IRSA).

\subsection*{White dwarf parameters}
\label{sec:WD}
The effective temperature and the surface gravity of the white dwarf component were estimated in a previous study as $T_1=36,965\pm100$\,K and $\log g_1=7.192\pm0.014$, respectively, based on an atmospheric fit to the original SPY UVES spectra from 2000 \cite{Koester_2009}. This is done by fitting the Balmer absorption lines using theoretical model spectra. In this fit the authors assumed a single white dwarf, but since both SPY epochs were taken, by chance, near orbital phase 0 (see below) when the irradiated day side of the companion is hidden, this assumption is valid. These parameters can be converted into a mass, a radius, and a cooling age using theoretical evolutionary tracks, by assuming a specific white-dwarf core composition: helium (He) or hybrid. The evolutionary tracks were computed using the Modules for Experiments in Stellar Astrophysics (\textsc{MESA}) code \cite{Paxton_2011, Paxton_2013, Paxton_2015, Paxton_2018, Paxton_2019}, similarly to Istrate et al. \cite{Istrate_2016} for He-core white dwarfs, and to Zenati et al. \cite{Zenati_2019} for hybrid-core white dwarfs (A.G.I., Justham, G.N., Grassitelli, Marchant, Romero, Klencki, Pols, Schwab, and Parsons, in preparation). The computed models include rotation and element diffusion. We created white dwarfs with various hydrogen-envelope masses, ranging from the canonical value resulting from binary evolution models (a few 10$^{-4}$\,\msun) down to 10$^{-10}$\,\msun, by artificially removing mass from the canonical white dwarf. We note that when we used canonical, stable mass-transfer, models (similar to the Althaus et al. \cite{Althaus_2013} models), the radius of the white dwarf was overestimated. For this reason we adopted the variable-envelope models here, in an attempt to mimic the results of a common envelope. The mass, radius, and cooling age of the white dwarf, interpolated using these two models, appear in Table~\ref{tab:param}.

In order to narrow down the possible parameter space, we estimated the white-dwarf radius using the GALEX ultraviolet measurements, in which the flux contribution from the companion should be negligible, as
\begin{equation}
    R_\textrm{WD} = \sqrt{\frac{f_{\lambda, \textrm{meas}}}{f_{\lambda, \textrm{theo}}}} d,
\end{equation}
where $f_{\lambda, \textrm{meas}}$ is the measured white-dwarf flux, $f_{\lambda, \textrm{theo}}$ is the extinction-corrected theoretical white-dwarf flux, and $d$ is the system's distance from Gaia DR3 (see Table~\ref{tab:param}). The estimated white-dwarf radius is $0.025954\pm0.00060$\,\rsun\ based on the GALEX FUV point, and $0.027137\pm0.00062$\,\rsun\ based on the GALEX NUV point. We thus adopted the full range, $0.0266\pm0.0012$\,\rsun, as the white-dwarf radius.

Another constraint for the theoretical models comes from the predicted surface abundance. As the white dwarf is relatively hot, gravitational settling timescale dictates a minimum white-dwarf mass for which, at this effective temperature, the surface is hydrogen dominated. The maximal allowed helium surface abundance was estimated by generating synthetic white-dwarf spectra with different helium abundances using the spectral synthesis program \textsc{Synspec} (version 50) \cite{Hubeny_2011}, based on a one-dimensional, horizontally homogeneous, plane-parallel, hydrostatic model atmosphere created with the \textsc{Tlusty} program (version 205) \cite{Hubeny_1988, Hubeny_1995, Hubeny_2017a}. The models were computed in local thermodynamic equilibrium (LTE), using the Tremblay tables \cite{Tremblay_2009} for hydrogen line broadening. The synthetic spectra were convolved with a $0.36$\,\AA-wide Gaussian, to mimic the UVES spectral resolution. We find that a helium surface abundance of $\sim 10^{-3}$ relative to hydrogen would have been detected in the UVES spectra. Since no helium lines are detected, the helium surface abundance must be $\lesssim 10^{-3}$ relative to hydrogen. We therefore excluded models where the hydrogen surface abundance is $\lesssim 1$.

\subsection*{Radial velocity analysis}
\subsubsection*{Radial velocity extraction}

As mentioned above, the only spectral features originating from the system that are detected in the UVES data are hydrogen Balmer lines---in absorption from the white dwarf, and in emission from the companion. The companion emission has a complex, `inverted', shape (see Fig.~\ref{fig:Halpha}a and Extended Data Fig.~\ref{fig:Halpha1D}) due to non-LTE (NLTE) effects \cite{Barman_2004}, as seen in other systems with low-mass irradiated companions (e.g. \cite{Parsons_2010}). The inverted line profile is seen in all of the Balmer lines. Since it has the best radial-velocity accuracy, we only use the H$\alpha$ line in our fit.

As the system rotates, the centres of the spectral lines of both of its components are shifted periodically in opposite directions due to Doppler effect. In order to extract the radial velocities of the white dwarf and the companion, we fitted a region of $\pm 1000$\,\kms\ around the position of the H$\alpha$ line, in each individual epoch, with a combination of a quadratic dependence of the flux on the velocity (fitting the wings of the full H$\alpha$ line profile) and three Gaussians---one in absorption, fitting the NLTE core of the H$\alpha$ line of the white dwarf, and a pair of Gaussians with inverted intensities sharing the same mean, fitting the inverted-core of the H$\alpha$ emission from the companion (see Extended Data Fig.~\ref{fig:Halpha1D}):
\begin{equation}
\label{eq:fit}
    \begin{split}
        I \left( v \right) = a_0 + a_1 v + a_2 v^2 - 
        & I_1 \exp{\left(-\frac{v - v_1}{2\sigma_1^2}\right)} +\\
        & I_\textrm{2, em} \exp{\left(-\frac{v - v_2 }{2\sigma_\textrm{2, em}^2}\right)} -
        I_\textrm{2, ab} \exp{\left(-\frac{v - v_2}{2\sigma_\textrm{2, ab}^2}\right)},
    \end{split}
\end{equation}
where $v_1$ and $v_2$ are the radial velocities of the white dwarf and the companion, respectively. $I_1$ and $\sigma_1$ are the intensity and width of the white-dwarf NLTE core absorption. $I_\textrm{2, em}$ and $I_\textrm{2, ab}$ are the intensities of the emission and absorption line components of the companion, respectively, while $\sigma_\textrm{2, em}$ and $\sigma_\textrm{2, ab}$ are the respective line component widths. All the parameters were fitted individually for each epoch. We note that the fitted Gaussian widths $\sigma_1$, $\sigma_\textrm{2, ab}$ and $\sigma_\textrm{2, em}$ varied by $\approx 30\%$, $70\%$, and $20\%$, respectively. This behaviour is known from similar systems (e.g. \cite{Maxted_1998, Parsons_2010}), and is likely caused by high optical depth and saturation effects in the lines.

Since the companion's emission is only visible when its irradiated day side is facing us, we first examined each epoch by eye and marked the epochs in which only the white-dwarf component is seen. We then fitted these epochs with a combination of the quadratic dependence and a single Gaussian, omitting the companion's contribution in equation~(\ref{eq:fit}).
The fit was performed using \textsc{scipy}'s \textsc{curve\_fit} bounded non-linear least squares Trust Region Reflective algorithm. The best-fitting line profiles are shown in Extended Data Fig.~\ref{fig:Halpha1D}. The radial velocity uncertainty was estimated based on the covariance matrix of each fit. Each epoch was assigned a barycentric timestamp at mid-exposure, and the velocities were shifted to the barycentric frame using \textsc{astropy}.

\subsubsection*{Orbital solution}
The radial velocity curves of the white dwarf and the companion were modelled using
\begin{equation}
    v_{1,2} = \gamma_{1,2} \pm K_{1,2} \sin \left[ 2 \pi \left( \phi - \phi_0 \right) \right],
\end{equation}
where $\gamma_{1,2}$ and $K_{1,2}$ are the systematic mean velocities and the radial-velocity semi-amplitudes of the white dwarf and companion, respectively, and $\phi_0$ is the initial orbital phase. Since the companion's emission originates from its irradiated side, we measure in fact the centre-of-light radial velocity, and not the centre-of-mass radial velocity. We denote the radial-velocity semi-amplitude of the companion's centre-of-light $K_\textrm{em}$, and correct it to the centre-of-mass frame below.

In order to probe the orbital period of the system over a wide range of values, we first examined the Lomb-Scargle periodogram of the radial velocity series of the white-dwarf component using \textsc{astropy}. The Lomb-Scargle periodogram computes the best-fit model parameters, $\vec{\theta}$, at a given frequency, $f$, for the model:
\begin{equation}
    y \left( t; f, \vec{\theta} \right) = \theta_0 + \sum_{n=1}^{\text{nterms}}\left[\theta_{2n-1}\sin(2\pi nft) + \theta_{2n}\cos(2\pi nft)\right].
\end{equation}
Assuming a circular orbit ($\textrm{nterms}=1$), and subtracting the weighted mean of the input data, $\delta$, before the fit, we can use the Lomb-Scargle fit parameters in order to estimate the radial-velocity semi-amplitude of the white dwarf,
\begin{equation}
    K_1 = \sqrt{\theta_1^2 + \theta_2^2},
\end{equation}
the initial orbital phase,
\begin{equation}
    \phi_0 = -\arctan\left( \frac{\theta_2}{\theta_1} \right),
\end{equation}
and the systematic mean velocity of the white dwarf,
\begin{equation}
    \gamma_1 = \theta_0 + \delta.
\end{equation}
The Lomb-Scargle periodogram of the radial velocity curve of the white dwarf is shown in Extended Data Fig.~\ref{fig:LS}. The best-fitting orbital period is $P=8340.3046\pm0.0075$\,s, and the model parameters are $K_1 \approx 52.7$\,\kms, $\phi_0 \approx 0.98$, and $\gamma_1 \approx 19.8$\,\kms.

We then used the Lomb-Scargle solution as an initial guess for a Markov chain Monte Carlo (MCMC) fit of the full radial-velocity data, including that of the companion. The fitting was performed using \textsc{emcee}, the \textsc{Python} implementation of the Affine Invariant MCMC Ensemble sampler \cite{Goodman_2010, ForemanMackey_2013}. The MCMC algorithm minimises the $\chi^2$ value of the fit over a six-dimensional parameter space defined by the orbital period ($P$), the initial orbital phase ($\phi_0$), the radial-velocity semi-amplitudes of the white dwarf ($K_1$) and the companion's emission ($K_\textrm{em}$), the mean radial velocity of the white dwarf ($\gamma_1$), and the difference between the mean velocities of the companion and the white dwarf ($\Delta \gamma \equiv \gamma_1 - \gamma_2$). The MCMC run included an ensemble of 25 `walkers' with 100,000 iterations each. The initial position of each walker was drawn from a Gaussian distribution around the initial guess, with a width of $10^{-10}$ for the orbital period, 0.01 for the initial orbital phase, and 0.1 for the rest of the fit parameters.
The auto-correlation lengths of the resulting MCMC chains ranged from 70 to 81 iterations. We thus discarded the first 161 iterations of each walker (`burn-in'), and kept every 34 iterations of the remaining walker chain (`thinning'). At the end of the process, each fit parameter had a final chain with a length of 73,400.
Fig.~\ref{fig:Halpha}b shows the radial velocity curve best fit, while Extended Data Fig.~\ref{fig:MCMC} shows the one- and two-dimensional projections of the posterior probability distributions of the fit parameters. The best-fitting parameters are given in Table~\ref{tab:param}.

We note that the relatively large uncertainty of $\Delta \gamma=11.4 \pm 1.7$\,\kms\ prevents us from estimating a meaningful secondary mass-radius ratio based on the gravitational redshift ($M_2/R_2=M_1/R_1 - \Delta\gamma c/G$). However, it is consistent within $0.9\sigma$ with the theoretical gravitational redshift of a He-core white dwarf ($9.86\pm0.14$\,\kms), and within $1.2\sigma$ with that of a hybrid-core white dwarf ($9.39\pm0.35$\,\kms), based on the white-dwarf parameters in Table~\ref{tab:param}.

\subsection*{Photometry analysis}
\label{sec:phot}

Modelling the light curves of a non-eclipsing system with an irradiated companion is a challenging task that depends on many poorly constrained highly degenerate parameters, and on the unknown details of the heat redistribution processes in the irradiated companion. We thus defer the light-curve modelling to future work, and focus instead on a comparison of the companion's day and night sides.
We have fitted the light curves with a simple sinusoidal model to guide the eye using \textsc{scipy}'s \textsc{curve\_fit} (Extended Data Fig.~\ref{fig:LC}). The Lomb-Scargle periodogram of each light curve, computed using \textsc{astropy}, is shown in Extended Data Fig.~\ref{fig:LC_LS}. The frequency of the highest peak in all of the light curves is consistent with the one of the radial-velocity curve (Extended Data Fig.~\ref{fig:LS}).

The calibrated light curves were phase-folded over the period and ephemeris obtained from the radial-velocity analysis (see Table~\ref{tab:param}), and binned into 50 bins by taking the median of each bin as the value, and 1.48 times the median absolute deviation divided by the square root of the number of data points in the bin as the error. The normalised phase-folded light curves are shown in Extended Data Fig.~\ref{fig:LC}. A clear irradiation effect is seen in the light curves, with no detected ellipsoidal modulation (expected at the $\sim1\%$ level) or eclipses (the expected eclipse duration is $\approx9$\,min, or about 6\% of the orbital period). The reflection contribution should be at a level of $\sim0.1\%$ \cite{Faigler_2011}.

The minimum (maximum) flux of the system was measured by taking the median flux $\pm0.05$ around orbital phase 0 (0.5), in each band. The error was calculated as 1.48 times the median absolute deviation of the flux divided by the square root of the number of data points. Given the rather sparse and noisy WISE \textit{W1}-band light curve, in this band we took the median flux $\pm0.1$ around orbital phases 0 and 0.5 as the minimum and maximum flux values, and 1.48 times the median absolute deviation of the minimal flux level as the error for both values. We then combined these extremum measurements in the $r'$, $i'$, $z$, and $W1$ bands with the archival GALEX FUV and NUV measurements (where the contribution from the irradiated companion is negligible), in order to estimate the companion's radius and night- and day-side effective temperatures. This was done by fitting the spectral energy distribution of the system with a combination of a white-dwarf model spectrum with a brown-dwarf model spectrum for the cooler night side, and with a black-body spectrum for the day side (Extended Data Fig.~\ref{fig:SED}). For the white dwarf we used a hydrogen-dominated DA model with an effective temperature of 37,000\,K and $\log g=7.2$ \cite{Levenhagen_2017}, and for the companion we used \textsc{BT-Dusty} models with effective temperatures ranging from 1,000 to 6,000\,K \cite{Allard_2011, Allard_2012}. We re-ran the fit using different surface gravity and metallicity values for the brown-dwarf models ($\log g$ of $5.0$ and $5.5$, and $[\textrm{M/H}]$ of $-1.0$, $-0.5$, and $0$). Models with $[\textrm{M/H}]=-1.0$ were available only for $\log g=5.5$ at this temperature range. All models were obtained from the Spanish Virtual Observatory (\url{http://svo.cab.inta-csic.es}).
Since the white-dwarf model truncates at a wavelength of $25,000$\,\AA, we have extrapolated it to $50,000$\,\AA\ assuming a Rayleigh-Jeans $\lambda^{-4}$ slope, where $\lambda$ is the wavelength.
The combined theoretical models were scaled using the estimated radii and the system's distance from Gaia DR3, and were reddened for extinction by Galactic dust (using \url{https://irsa.ipac.caltech.edu/applications/DUST/})\cite{Schlafly_2011}. We then fitted the observed spectral energy distribution with the band-integrated theoretical flux using the \textsc{emcee} package. The MCMC algorithm minimises the $\chi^2$ value of the fit over a four-dimensional parameter space defined by the companion's night- and day-side effective temperatures ($T_2^\text{night}$ and $T_2^\text{day}$, respectively), the companion's radius ($R_2$), and the fraction of night/day-side contamination due to the system's inclination and the companion's heat distribution ($f_\text{cont}$). The MCMC run included an ensemble of 25 `walkers' with 40,000 iterations each. The initial position of each walker was drawn from a Gaussian distribution with a width of 0.3 around the initial guess ($T_2^\text{night}=3000$\,K, $T_2^\text{day}=6000$\,K, $R_2=0.08$\,\rsun, $f_\text{cont}=0.2$).
The resulting minimal $\chi^2$ value was slightly lower for the fit that uses brown-dwarf models with $\log g=5.5$ (although insignificantly, by $\sim 0.004$). Among the $\log g=5.5$ model fits, models with $[\textrm{M/H}]=-0.5$ had slightly lower minimal $\chi^2$ values assuming a He-core white dwarf (by $\sim 0.012$ compared to $[\textrm{M/H}]=0$ and by $\sim 0.004$ compared to $[\textrm{M/H}]=-1.0$), or $[\textrm{M/H}]=-1.0$ assuming a hybrid-core white dwarf (by $\sim 0.003$ compared to $[\textrm{M/H}]=-0.5$ and by $\sim 0.004$ compared to $[\textrm{M/H}]=0$). We have thus adopted the results using values of $\log g=5.5$ and $[\textrm{M/H}]=-0.5$ for a He-core white dwarf, and $\log g=5$ and $[\textrm{M/H}]=-1.0$ for a hybrid-core white dwarf.
The auto-correlation lengths of the resulting MCMC chains ranged from 54 to 69 iterations. We thus discarded the first 125 (137) iterations of each walker (`burn-in'), and kept every 26 (27) iterations of the remaining walker chain (`thinning') for the fit assuming a He-core (hybrid) white-dwarf radius. At the end of the process, each fit parameter had a final chain with a length of 38,325 (36,900).
The best-fit models are plotted in Fig.~\ref{fig:SED}, and listed in Table~\ref{tab:param}. Extended Data Figs.~\ref{fig:MCMC_SED_He} and \ref{fig:MCMC_SED_Hybrid} show the one- and two-dimensional projections of the posterior probability distributions of the fit parameters.

\subsection*{Correcting $K_\textrm{em}$ for the centre of mass}
\label{sec:K2}

As mentioned above, the companion's emission originates from the surface of its irradiated side. The radial velocities measured from the emission line thus impose a lower limit on the centre-of-mass radial velocities \cite{Parsons_2010}. The radial velocity semi-amplitude of the centre of mass, $K_2$, can be estimated by
\begin{equation}\label{eq:K2}
    K_2 = \frac{K_\textrm{em}}{1 - f \left(1 + q \right) \frac{R_2}{a}},
\end{equation}
where $q=M_2/M_1$ is the binary mass ratio, $R_2$ is the radius of the companion, $a$ is the orbital separation, and $0 \leq f \leq 1$ is a constant that depends upon the location of the centre of light \cite{Parsons_2012}. For an optically thick line such as the H$\alpha$ line, we can assume $f\approx0.5$ (as demonstrated by \cite{Parsons_2010, Parsons_2012}).

The orbital separation can be calculated using Kepler's law,
\begin{equation}\label{eq:a}
    a = \left( \frac{P^2}{4\pi^2}G\left( M_1 + M_2 \right) \right)^{1/3},
\end{equation}
where $P$ is the orbital period, $G$ is the gravitational constant, $M_1$ is the white-dwarf mass, and $M_2=qM_1$ is the mass of the companion.
Since $q=M_2/M_1=K_1/K_2$, there is only a single $q$ value that is consistent with both equations~(\ref{eq:K2}) and (\ref{eq:a}), given the measured values of $K_\textrm{em}$, $P$, and $R_2$, and the assumed values of $M_1$ and $f$. Table~\ref{tab:param} lists the derived values of $q$ and $K_2$ for each white-dwarf core composition.

We then calculate the orbital inclination, $i$:
\begin{equation}\label{eq:i}
    i = \arcsin \left[ \left( \frac{P}{2\pi G M_1} \left(K_1 + K_2 \right)^2 K_2 \right)^{\frac{1}{3}} \right],
\end{equation}
assuming a circular orbit. The implied possible orbital inclination range is listed in Table~\ref{tab:param}.

\subsection*{Equilibrium temperature}
The `equilibrium' temperature of the irradiated companion (neglecting its intrinsic luminosity and albedo, and assuming it is in thermal equilibrium with the external irradiation) is listed in Table~\ref{tab:param}, for each white-dwarf core composition. It is defined as
\begin{equation}
    T_\textrm{eq} \equiv T_\textrm{1} \sqrt{ \frac{R_\textrm{1}}{2a}},
\end{equation}
where $T_\textrm{1}$ and $R_\textrm{1}$ are the effective temperature and radius of the white dwarf, and $a$ is the orbital separation \cite{Arras_2006}.

\subsection*{Near-infrared spectroscopy with FLAMINGOS-2}
We obtained a pair of low-resolution near-infrared spectra, around orbital phases 0 and 0.35, on 2022 June 9 using the FLAMINGOS-2 spectrograph \cite{Eikenberry_2004} on Gemini South in Cerro Pach\'{o}n, Chile. The observations were carried out using the HK grism, HK filter, and a 0.36\,arcsec slit, covering the \textit{H}- and \textit{K}-band region ($\approx13,000-21,500$\,\AA) with a spectral resolution of $R\approx 900$. Each spectrum was composed of five 2-min exposures. The telescope was nodded along the slit between the exposures to facilitate the sky subtraction.
The data of the target and of the telluric standard HD\,225187 were reduced and the raw count spectra were extracted using the \textsc{gemini iraf} package version 1.14, following the Gemini F2 Longslit Tutorial (\url{https://gemini-iraf-flamingos-2-cookbook.readthedocs.io/en/latest/Tutorial_Longslit.html}).

We then used the \textsc{sparta} \textsc{python} package (\url{https://github.com/SPARTA-dev/SPARTA}) in order to retrieve and broaden to the FLAMINGOS-2 spectral resolution a \textsc{phoenix} model spectrum of the telluric star. We normalised it by dividing it by its continuum shape (obtained by interpolating over the line-free regions in the spectrum), and applied to it the expected Doppler shift at the time of the observations. We manually scaled the normalised model spectrum so that its absorption lines agree with those in the raw count spectrum of the telluric star. Finally, we divided the telluric star's raw spectrum by the scaled model spectrum to remove the star's intrinsic absorption lines from the observed telluric spectrum. We then divided the raw spectrum of \obj\ by the telluric spectrum, taking the different exposure times into account, to obtain the relative count spectrum of \obj. We calibrated the flux using the telluric star's archival $H$-band magnitude, assuming a black-body model. Finally, we binned the result by taking the median of every four data points (Extended Data Fig.~\ref{fig:F2}).

\subsection*{Formation history}
We estimate the white dwarf progenitor mass, $M_\textrm{MS}$, as \cite{Nelemans_1998}
\begin{equation}
    M_\textrm{MS} \approx \frac{1}{2}M_1 \left(1 + \sqrt{1 + \frac{2 \alpha \lambda R_\textrm{RG}  M_2}{M_1 a_0}} \right),
\end{equation}
where $M_1$ is the white dwarf mass, $M_2$ is the mass of the companion, and we have assumed $M_2 \ll M_\textrm{MS} - M_1$. $R_\textrm{RG} $ is the radius of the progenitor red giant in the beginning of the spiral-in phase. In the case of a He-core white dwarf, it can be approximated as \cite{Iben_1984}
\begin{equation}
    R_\textrm{RG} \approx 10^{3.5} \left( \frac{M_1}{\textrm{\msun}} \right)^4\,\textrm{\rsun},
\end{equation}
corresponding to $R_\textrm{RG, He} \approx 97$\,\rsun.
After the envelope ejection, the orbital separation shrinks with time due to gravitational-wave emission. The orbital separation immediately after the envelope ejection, $a_0$, is estimated as \cite{Maoz_2012}
\begin{equation}
    a_0 = \left[a^4 + \frac{256}{5}\frac{G^3}{c^5} M_1 M_2 \left(M_1 + M_2\right) \Delta t \right]^\frac{1}{4},
\end{equation}
where $a$ is the present-day orbital separation, and $\Delta t$ is the time that has passed since the envelope ejection, approximated as the white-dwarf cooling age, $t_1$. Given the young cooling age of the white dwarf ($\sim 1$\,Myr), the orbital separation has changed by merely $\sim 0.01\,\%$.
$\alpha \equiv \Delta E_\textrm{bind} / \Delta E_\textrm{orb}$ is a parameter describing the envelope ejection efficiency, and $\lambda < 1$ is a weighting factor that depends on the structure of the red giant. For $\lambda=0.5$ and $\alpha$ ranging between 0.5 and 4 \cite{Tauris_1996, PortegiesZwart_1998}, we get a white dwarf progenitor mass ranging between $\approx 1-2.4$\,\msun\ for a He-core white dwarf.

The small radius of the companion indicates an age of at least a few Gyr (Fig.~\ref{fig:MRR})\cite{Phillips_2020}. On the other hand, the white-dwarf cooling age---i.e. the time that has passed since it lost its envelope---is $\sim 1$\,Myr. This suggests that the companion was not significantly heated during the common-envelope phase, indicating that the internal thermodynamic energy of the envelope did not contribute much to the envelope ejection ($\alpha \sim 1$). Assuming the full energy required to unbind the envelope came from orbital sources, the progenitor of a He-core white dwarf could have been quite a low-mass star of $\sim 1.3$\,\msun.

The critical mass above which the companion does not evaporate during the envelope ejection is \cite{Nelemans_1998}
\begin{equation}
    m_\textrm{crit} = 10 \left[ \left(\frac{M_\textrm{MS} - M_1}{M_1}\right) \left(\frac{M_\textrm{MS}}{\textrm{\msun}}\right) \left(\frac{R_\textrm{RG}}{100\,\textrm{\rsun}}\right)\right]^{0.46}\,\textrm{M}_\textrm{Jup},
\end{equation}
and ranges between $\approx 0.01-0.03$\,\msun\ for a He-core white dwarf---well below the mass of the companion.

Hybrid-core white dwarfs, on the other hand, are the descendants of more massive and compact system, with a factor $\gtrsim 5$ larger binding energies (e.g. \cite{Hu_2007}). To estimate the envelope binding energy in the hybrid scenario, we modelled a hybrid progenitor with a mass of $2.3$\,\msun\, and a He-core progenitor with a mass of $1.3$\,\msun, when both reached a He-core of 0.4\,\msun. At this stage, we find that the binding energy of the hybrid progenitor is about 26 times larger than that of the He-core progenitor. For the He-core progenitor we find $\lambda_\textrm{He} \approx 0.7$ and $\alpha_\textrm{He} \approx 1.1$, while for the hybrid progenitor we find $\lambda_\textrm{Hybrid} \approx 0.9$ and $\alpha_\textrm{Hybrid} \approx 31$. This would require unbinding the envelope with a much higher efficiency in order for the companion to survive and get to the observed close orbit, and might argue against a hybrid nature of the white dwarf. However, since the exact physical processes governing the common envelope evolution are unknown, a hybrid-core white dwarf cannot be excluded.

\subsection*{The white dwarf + brown dwarf population}
To date, only twelve white dwarf + brown dwarf systems are known \cite{Farihi_2004, Burleigh_2006, Maxted_2006, Casewell_2009, Burleigh_2011, Casewell_2012, Beuermann_2013, Steele_2013, Littlefair_2014, Farihi_2017, Parsons_2017, Casewell_2018, Casewell_2020, vanRoestel_2021, RebassaMansergas_2022}. This makes \obj\ the thirteenth known such system (assuming the companion is a brown dwarf), with the hottest irradiated companion (see Fig.~\ref{fig:WDBD}).
There is an additional candidate white dwarf + brown dwarf system SDSS\,J1231+0041 \cite{Parsons_2017}, that somewhat resembles \obj\ (with an equilibrium temperature $\approx 400$\,K cooler). However at a distance of $\approx 1,500$\,pc and an apparent magnitude of $G=20.35$ (compared to $G=16.10$ of \obj), it is difficult to obtain time-resolved spectroscopy for this system and to confirm the nature of the heated companion. Given this observational challenge, this system cannot serve as a useful ultra-hot Jupiter analogue.

\obj\ was identified as a binary candidate out of a sub-sample of 439 white dwarfs from the SPY survey \cite{Maoz_2017}. Incidentally, WD\,0137$-$349, the first confirmed post-common-envelope white dwarf + brown dwarf binary, was also discovered by an early analysis of the SPY data \cite{Maxted_2006, Burleigh_2006}, which included $\sim 800$ white dwarfs. Current lower limits on the white dwarf + brown dwarf binary fraction are $f\geq 0.5\pm0.3$\% \cite{Steele_2011}, and $f>0.8-2\%$ \cite{Girven_2011}. Given that these binary fraction estimates were for all orbital separations, while the radial-velocity changes detectable by SPY limit the white dwarf + brown dwarf systems that it can find to $\lesssim 0.1$\,AU \cite{Maoz_2017}, the observed incidence is consistent with both of these previous estimates.

\backmatter

\bigskip

\bmhead{Data Availability}

The UVES spectroscopic data are available through the ESO archive facility (\url{http://archive.eso.org/cms.html}) under programme IDs 165.H-0588(A), 0103.D-0731(A), and 105.20NQ.001.

The FLAMINGOS-2 spectroscopic data are available through the Gemini Observatory archive (\url{https://archive.gemini.edu}) under program ID GS-2022A-FT-108.

The LCOGT photometric data are available at the LCOGT science archive (\url{https://archive.lco.global}) under program IDs TAU2021B-004 and TAU2022B-004.

The TESS photometric data are publicly available from the Mikulski Archive for Space Telescopes (MAST; 
\url{https://mast.stsci.edu}).

The WISE photometric data are publicly available from the Infrared Processing and Analysis Center (IPAC) Infrared Science Archive (IRSA; \url{https://irsa.ipac.caltech.edu/}).

The white-dwarf theoretical evolutionary tracks used in the analysis will be published in a future publication led by A.G.I., and are available upon request from the corresponding author.

The source data of Fig.~\ref{fig:Halpha} (radial velocity) and Extended Data Figs.~\ref{fig:LC} (light curves) and \ref{fig:F2} (infrared spectra) are published alongside this manuscript.

\bmhead{Code Availability}
We used various publicly available software packages in our analysis, all of them are mentioned in the relevant parts of the paper.

\bmhead{Acknowledgments}
We thank Sahar Shahaf for useful discussions, Jason Spyromilio for comments on the observing proposals and manuscript, John Pritchard from ESO User Support for assistance with the observation planning, and Avraham Binnenfeld for help in verifying the orbital period.

This work was supported by a Benoziyo-prize postdoctoral fellowship (N.H.).
This work was supported by a grant from the European Research Council (ERC) under the European Union's FP7 Programme, Grant No. 833031 (D.M.).
A.G.I. acknowledges support from the Netherlands Organisation for Scientific Research (NWO).
C.B. acknowledges support from the National Science Foundation grant AST-1909022.
E.B. acknowledges support from the Science and Technology Facilities Council (STFC) grant no. ST/S000623/1.
B.T.G. acknowledges support from the UK's Science and Technology Facilities Council (STFC), grant ST/T000406/1. This project has received funding from the European Research Council (ERC) under the European Union’s Horizon 2020 research and innovation programme (Grant agreement No. 101020057).
A.R.M. acknowledges support from the Spanish MINECO grant PID2020-117252GB-I00 and from the AGAUR/Generalitat de Catalunya grant SGR-386/2021.
F.M. acknowledges support from the INAF Large Grant ``Dual and binary supermassive black holes in the multi-messenger era: from galaxy mergers to gravitational waves'' (Bando Ricerca Fondamentale INAF 2022), from the INAF project ``VLT-MOONS'' CRAM 1.05.03.07.

Based on observations collected at the European Southern Observatory under ESO programmes 165.H-0588(A), 0103.D-0731(A), and 105.20NQ.001. This research has made use of the services of the ESO Science Archive Facility.
This work makes use of observations from the Las Cumbres Observatory global telescope network under program TAU2021B-004.
Based on observations obtained at the international Gemini Observatory, a program of NSF’s NOIRLab, which is managed by the Association of Universities for Research in Astronomy (AURA) under a cooperative agreement with the National Science Foundation on behalf of the Gemini Observatory partnership: the National Science Foundation (United States), National Research Council (Canada), Agencia Nacional de Investigaci\'{o}n y Desarrollo (Chile), Ministerio de Ciencia, Tecnolog\'{i}a e Innovaci\'{o}n (Argentina), Minist\'{e}rio da Ci\^{e}ncia, Tecnologia, Inova\c{c}\~{o}es e Comunica\c{c}\~{o}es (Brazil), and Korea Astronomy and Space Science Institute (Republic of Korea).
This work was enabled by observations made from the Gemini North telescope, located within the Maunakea Science Reserve and adjacent to the summit of Maunakea. We are grateful for the privilege of observing the Universe from a place that is unique in both its astronomical quality and its cultural significance.
This paper includes data collected by the TESS mission, which are publicly available from the Mikulski Archive for Space Telescopes (MAST). Funding for the TESS mission is provided by the NASA's Science Mission Directorate.
This research has made use of the VizieR catalogue access tool, CDS, Strasbourg, France.
This research has made use of the Spanish Virtual Observatory (\url{http://svo.cab.inta-csic.es}) supported from Ministerio de Ciencia e Innovaci\'{o}n through grant PID2020-112949GB-I00.
This publication makes use of data products from the Wide-field Infrared Survey Explorer, which is a joint project of the University of California, Los Angeles, and the Jet Propulsion Laboratory/California Institute of Technology, funded by the National Aeronautics and Space Administration.
This work has made use of data from the European Space Agency (ESA) mission Gaia (\url{https://www.cosmos.esa.int/gaia}), processed by the Gaia Data Processing and Analysis Consortium (DPAC; \url{https://www.cosmos.esa.int/web/gaia/dpac/consortium}). Funding for the DPAC has been provided by national institutions, in particular the institutions participating in the Gaia Multilateral Agreement.
This research has made use of the NASA Exoplanet Archive, which is operated by the California Institute of Technology, under contract with the National Aeronautics and Space Administration under the Exoplanet Exploration Program.
This research has made use of the \textsc{Python} package \textsc{GaiaXPy} (\url{https://gaia-dpci.github.io/GaiaXPy-website/}, \url{https://doi.org/10.5281/zenodo.7374213}), developed and maintained by members of the Gaia Data Processing and Analysis Consortium (DPAC), and in particular, Coordination Unit 5 (CU5), and the Data Processing Centre located at the Institute of Astronomy, Cambridge, UK (DPCI).
This research made use of \textsc{astropy} (\url{http://www.astropy.org}), a community-developed core \textsc{python} package for Astronomy \cite{Astropy_2013, Astropy_2018}, \textsc{corner} \cite{ForemanMackey_2016}, \textsc{emcee} \cite{ForemanMackey_2013}, \textsc{lightkurve} \cite{Lightkurve_2018}, \textsc{matplotlib} \cite{Hunter_2007}, \textsc{numpy} \cite{Numpy_2006, Numpy_2011}, \textsc{scipy} \cite{Virtanen_2020}, \textsc{sparta} (\url{https://github.com/SPARTA-dev/SPARTA}), \textsc{stsynphot} \cite{Stsynphot_2020}, \textsc{synphot} \cite{Synphot_2018}, and \textsc{uncertainties} (\url{http://pythonhosted.org/uncertainties/}), a \textsc{python} package for calculations with uncertainties by Eric O. Lebigot.

\bmhead{Author contributions}
N.H. led the observational follow-up effort, analysed the data, and wrote the majority of this manuscript.
D.M. and N.H. have analysed the original SPY survey data, and flagged this object as a potential binary system.
A.G.I. generated and fitted the helium- and hybrid-core white-dwarf models.
S.W.J. was the principal investigator of the Gemini follow-up programme.
B.L., T.R.M., and G.N. were part of the team of the original SPY programme.
All of the authors applied for spectroscopic follow-up telescope time, contributed to the discussion, and commented on the manuscript.

\bmhead{Competing interests}
The authors declare no competing interests.

\begin{appendices}


\renewcommand{\figurename}{Extended Data Fig.}
\renewcommand\thefigure{\arabic{figure}}
\setcounter{figure}{0}

\begin{figure}[h]
    \centering
    \includegraphics[width=0.6\textwidth]{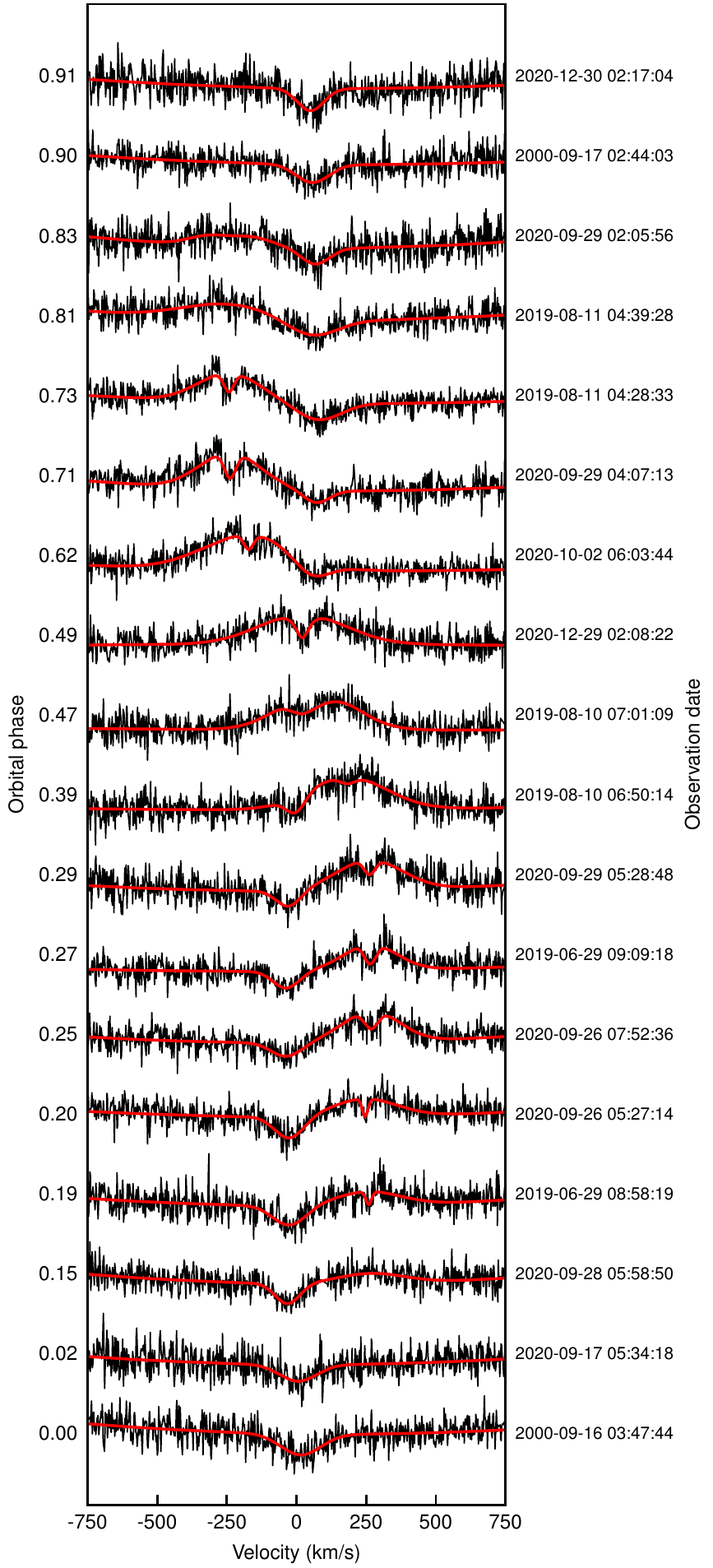}
    \caption{\textbf{H$\alpha$ line profile of \obj\ as a function of orbital phase.} H$\alpha$ line profile (black) and fit (red) of all the UVES epochs, vertically shifted for visual clarity, and sorted by orbital phase from bottom to top. The absorption line of the white-dwarf component is seen in all the epochs, while the inverted-core emission from the companion disappears when its night side is facing us.}
    \label{fig:Halpha1D}
\end{figure}

\begin{figure}[h]
    \centering
    \includegraphics[width=\textwidth]{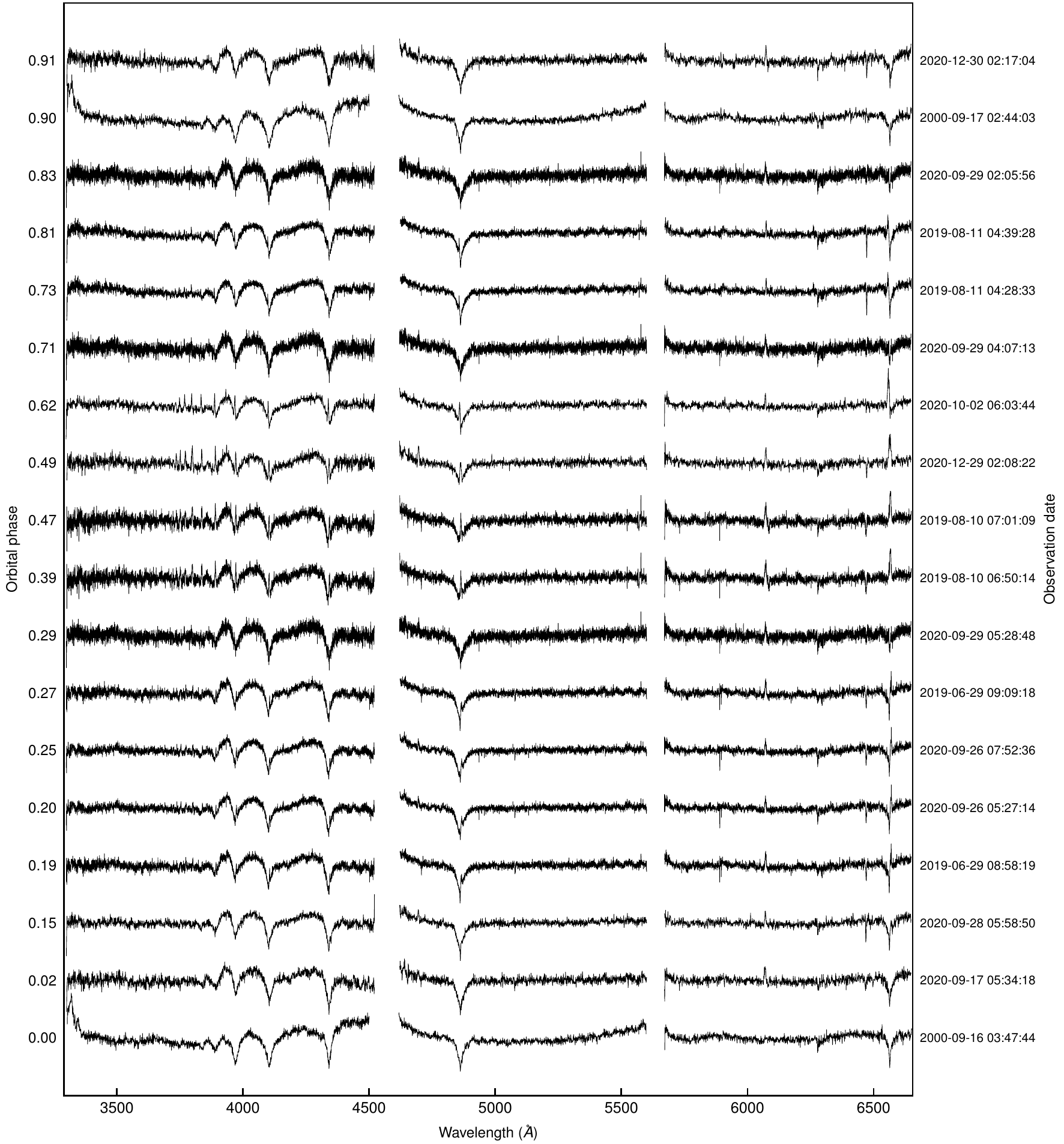}
    \caption{\textbf{Binned and normalised spectra of all the UVES epochs}, vertically shifted for visual clarity, and sorted by orbital phase from bottom to top. The Balmer line absorption of the white dwarf is seen throughout the orbital phase, while the companion's Balmer line emission is visible between phases 0.19 and 0.81. Other spectral lines seen in the spectra are of either telluric or interstellar origin, with fixed radial velocities with respect to the system's components.}
    \label{fig:FullSpectra}
\end{figure}

\begin{figure}[h]
    \centering
    \includegraphics[width=0.6\textwidth]{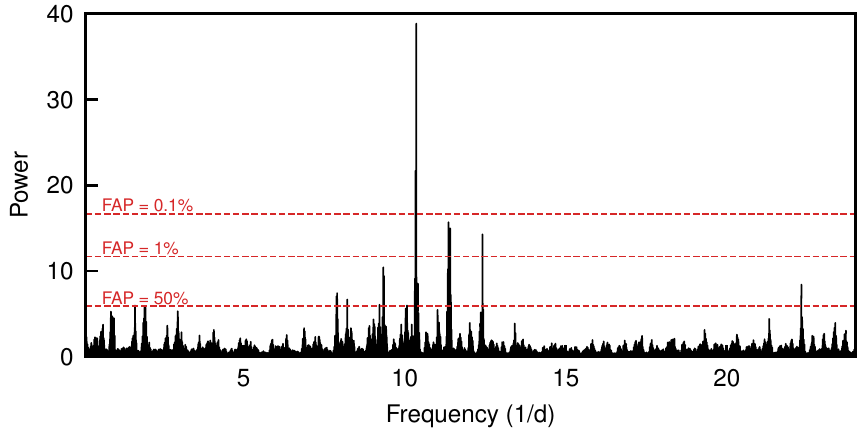}
    \caption{\textbf{Lomb-Scargle periodogram of the radial-velocity curve of the white-dwarf component.} The false-alarm probability (FAP) levels of 0.1, 1, and 50\% are marked with the red dashed lines.}
    \label{fig:LS}
\end{figure}

\begin{figure*}[h]
    \centering
    \includegraphics[width=\textwidth]{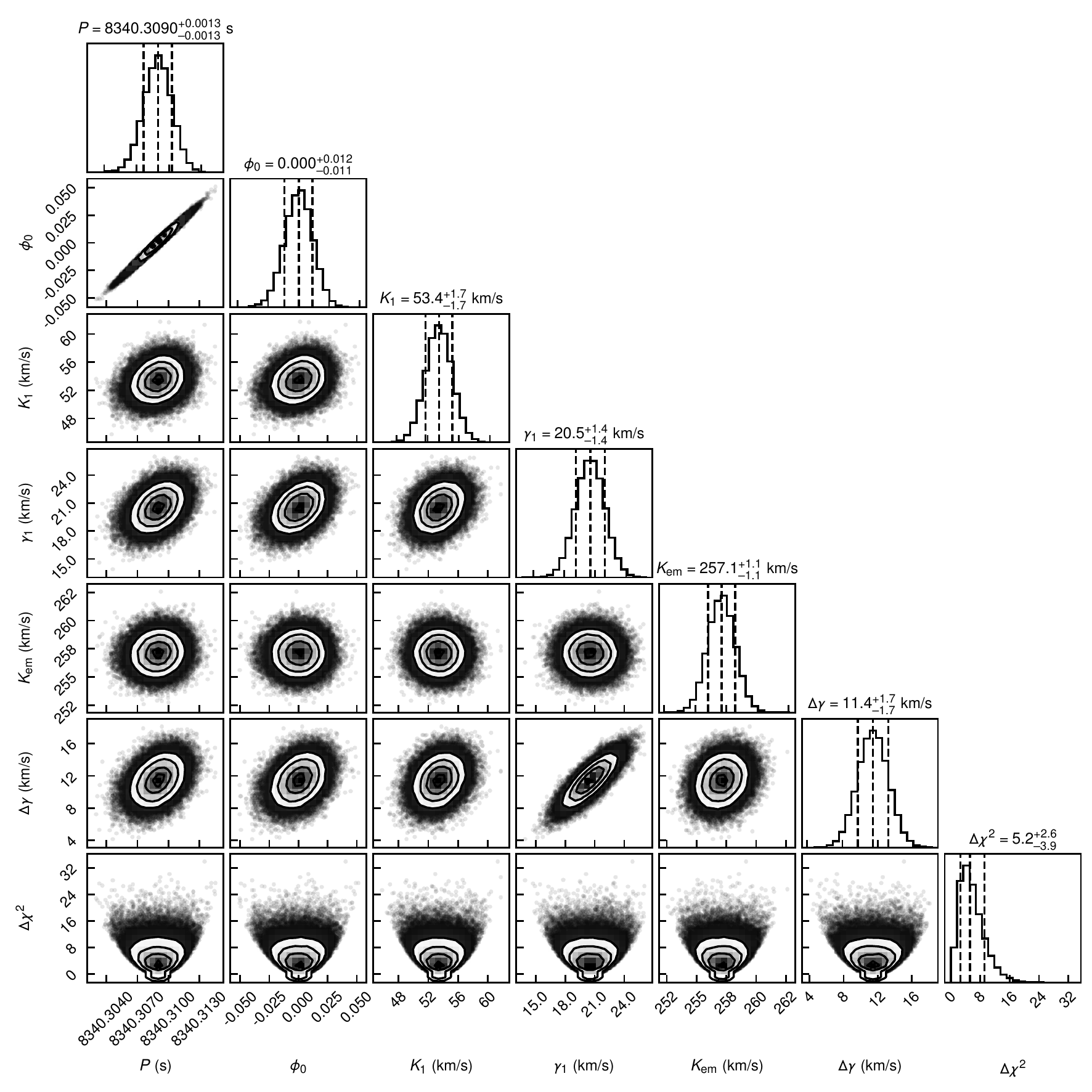}
    \caption{\textbf{One- and two-dimensional projections of the posterior probability distributions of the MCMC fit parameters for the radial-velocity curves.} The vertical dashed lines mark the median value and its $1\sigma$ uncertainty.}
    \label{fig:MCMC}
\end{figure*}

\begin{figure}[h]
    \centering
    \includegraphics[width=\textwidth]{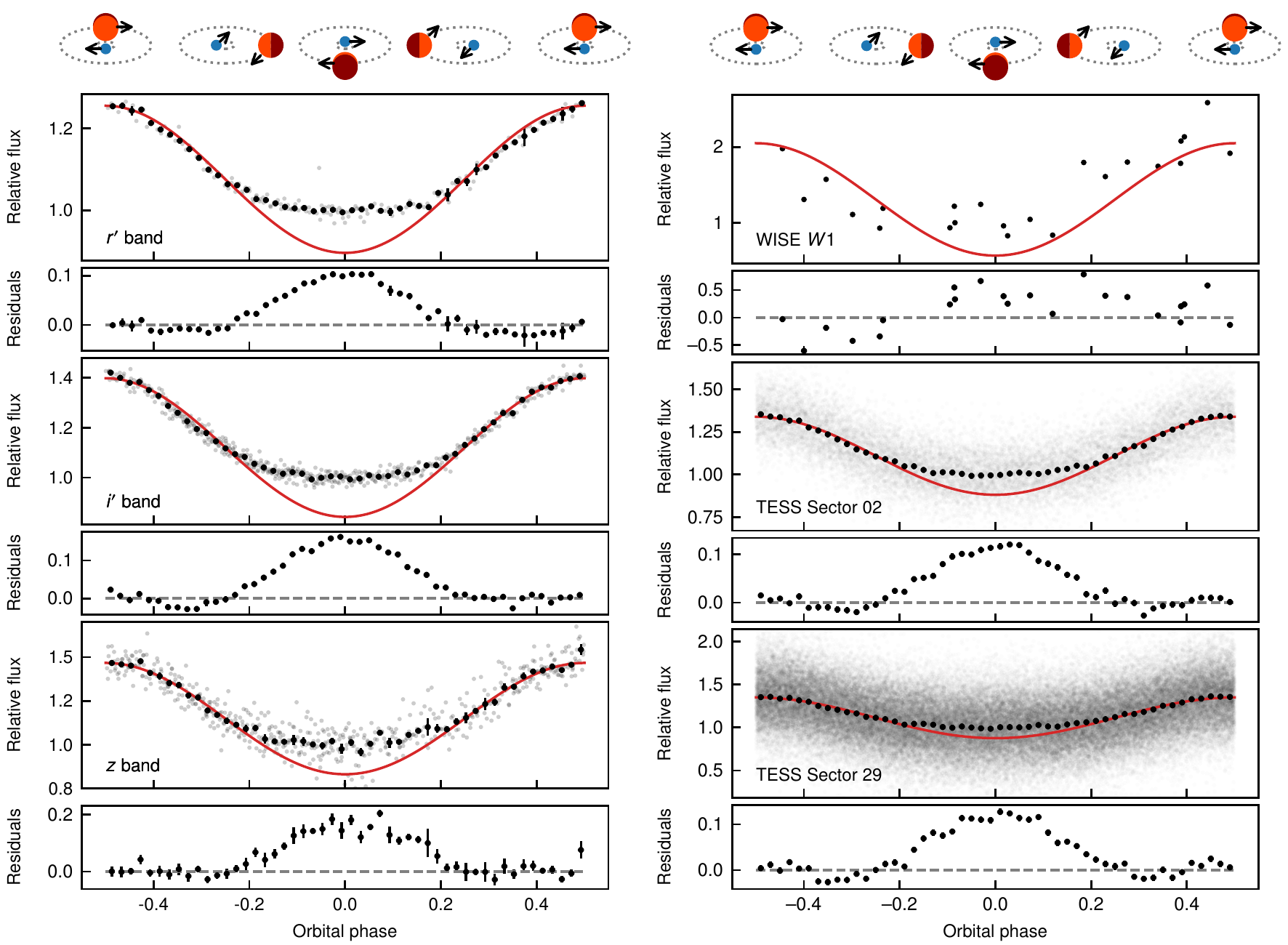}
    \caption{\textbf{Folded light curves of \obj.} Normalised (grey dots) and binned (black error bars) light curves of the \obj\ system from LCOGT (\textit{left}), WISE W1-band (\textit{top right}; unbinned), and TESS (\textit{middle and bottom right}), phase-folded over the orbital period ($P=8340.9090$\,s). No phase shift is seen between the various bands. The orbital period matches the one obtained from the spectroscopy.
    The error bars of the binned light curves show 1.48 times the median absolute deviation of the flux divided by the square root of the
    number of data points in each bin.
    A sine function fitted to orbital phases $\mid \phi \mid >0.2$ is plotted in red. The residual plot for each model is shown in the sub-panel below each light curve. The illustrations on \textit{top} demonstrate the system's configuration at each orbital phase. The flat bottom corresponds to the companion's night side.}
    \label{fig:LC}
\end{figure}

\begin{figure}[h]
    \centering
    \includegraphics[width=0.6\textwidth]{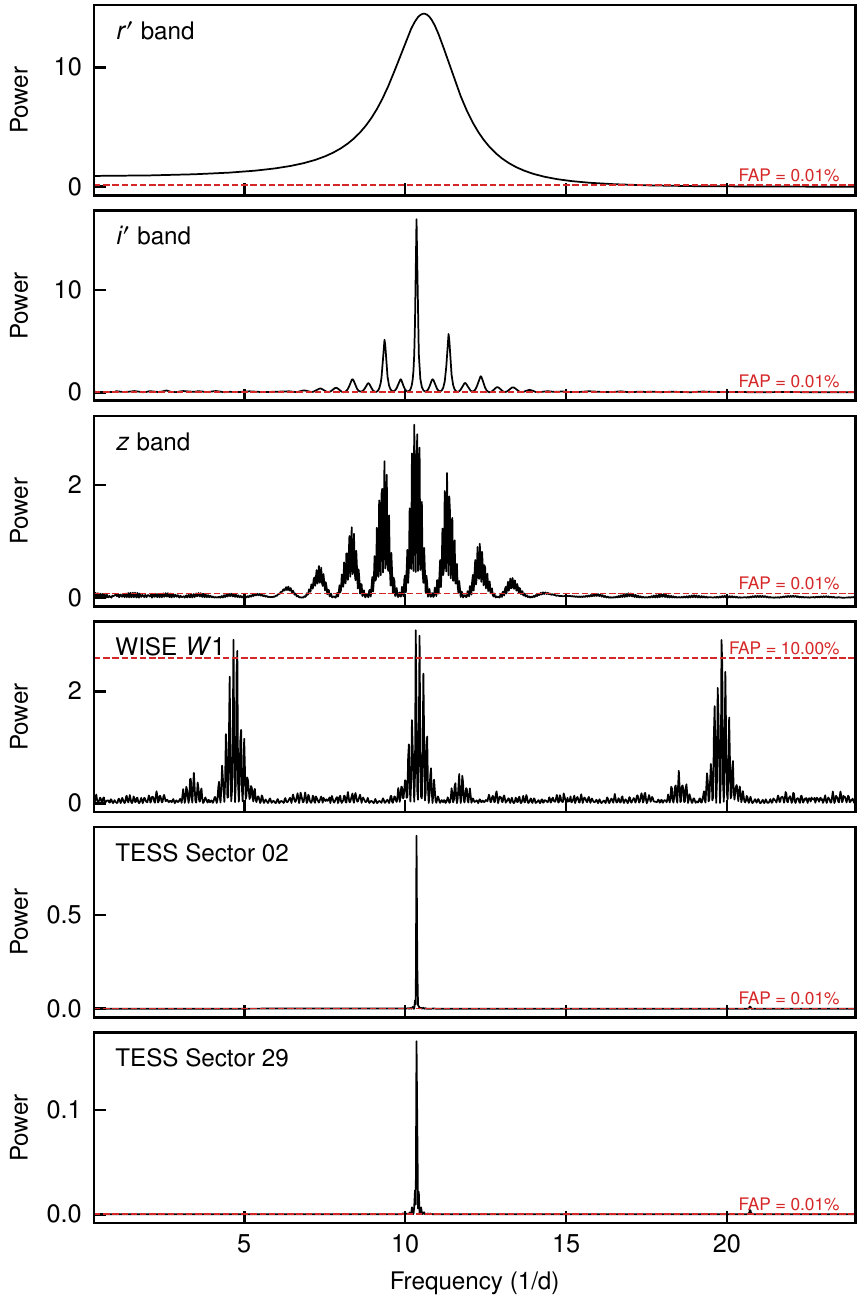}
    \caption{\textbf{Lomb-Scargle periodogram of the various light curves} (see Extended Data Fig.~\ref{fig:LC}). The false-alarm probability (FAP) level of 0.01\% (or 10\% for the WISE W1 band) is marked with a red dashed line. The frequency of the highest peak in all of the light curves is consistent with the one of the radial-velocity curve (Extended Data Fig.~\ref{fig:LS}).}
    \label{fig:LC_LS}
\end{figure}

\begin{figure*}[h]
    \centering
    \includegraphics[width=\textwidth]{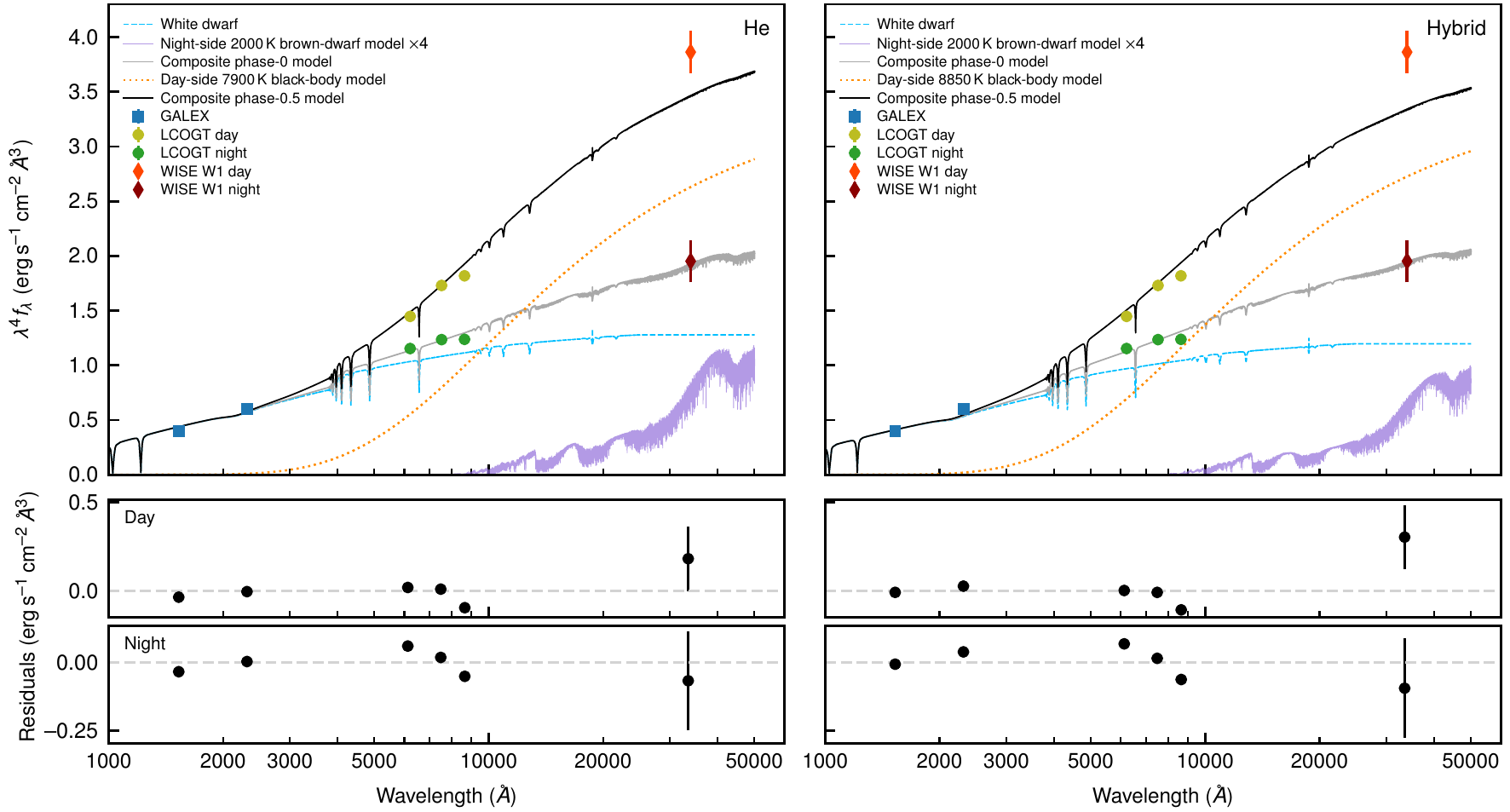}
    \caption{\textbf{Observed spectral energy distribution for \obj}, compared to the best-fitting composite theoretical model spectra of a white dwarf and a black body/brown dwarf. The archival GALEX ultraviolet photometry, where the contribution from the companion is negligible, appears as blue square-shaped error bars.
    Minimal/maximal photometric values in different bands, extracted from the light curves, appear as green-shades circle-shaped error bars for LCOGT's $r'$, $i'$, and $z$ bands, and as red-shades diamond-shaped error bars for the WISE W1 band.
    A theoretical model spectrum of a hydrogen-dominated white dwarf with an effective temperature of $37,000$\,K and a surface gravity $\log g=7.2$ \cite{Levenhagen_2017} is shown in dashed light blue. The best-fitting brown-dwarf (\cite{Allard_2011, Allard_2012}; for the night side, with $\textrm{[M/H]}=-0.5$ (He) or $\textrm{[M/H]}=-1.0$ (hybrid) and $\log g=5.5$) and black-body (for the day side) models are plotted in solid purple and dotted orange, respectively. The theoretical spectra were scaled using the system's distance measured by the Gaia mission, and the estimated component radii (\textit{left}: assuming a helium-core white dwarf (He), \textit{right}: assuming a `hybrid' carbon-oxygen core white dwarf with a thick helium envelope).
    The brown-dwarf model is shown multiplied by a factor of 4, to fit the displayed range.
    The composite model of the system at orbital phase 0 (0.5) is plotted in solid dark grey (black).
    The units shown on the $y$ axis are the flux per wavelength, $\lambda$, multiplied by $\lambda^4$, for visual clarity.
    The bottom panels show the residuals of the day-side (\textit{middle}) and the night-side (\textit{bottom}) fits. The error bars in the residual plots show the standard deviation and take into account both the photometric and the model uncertainties.}
    \label{fig:SED}
\end{figure*}

\begin{figure*}[h]
    \centering
    \includegraphics[width=\textwidth]{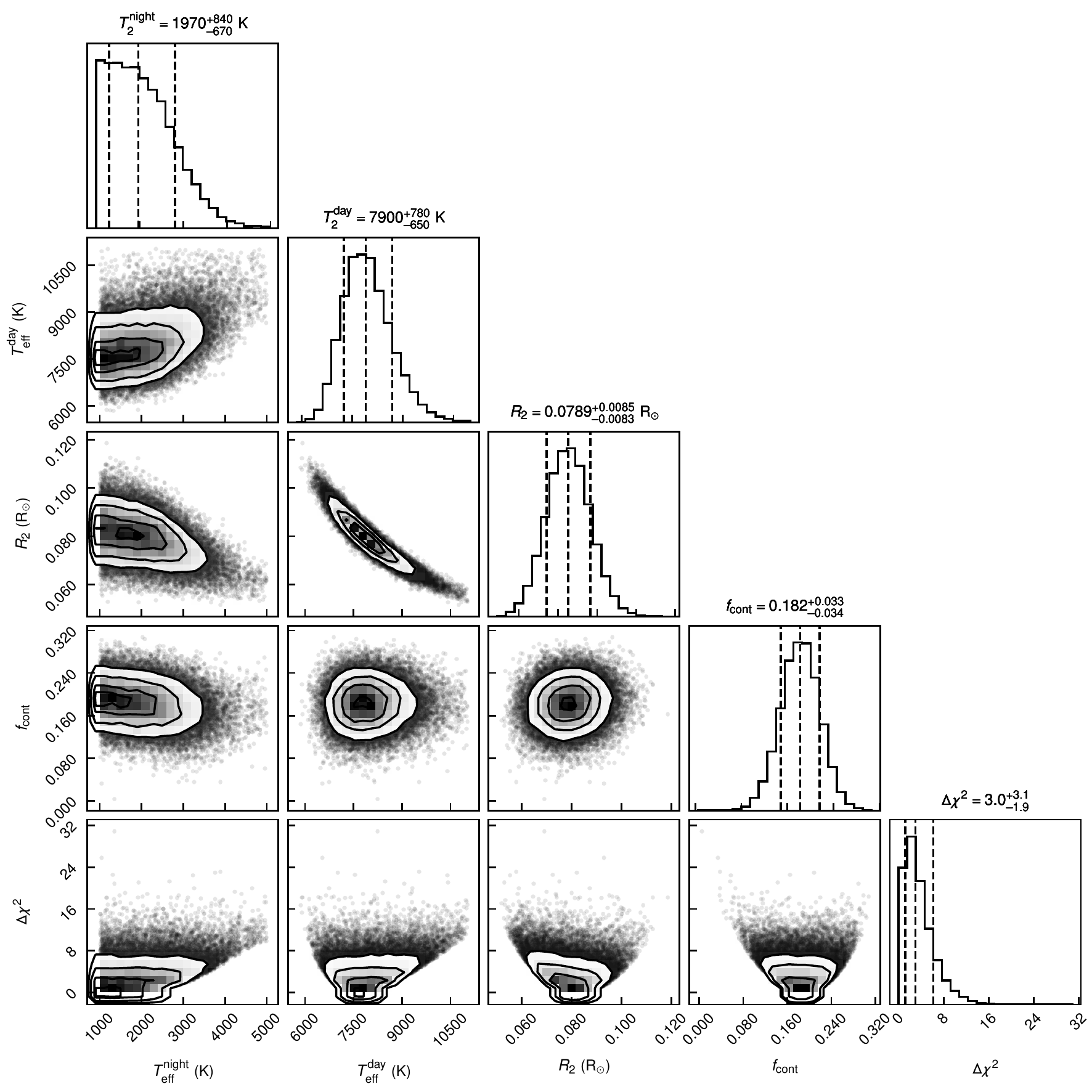}
    \caption{\textbf{One- and two-dimensional projections of the posterior probability distributions of the MCMC fit parameters for the spectral energy distribution, assuming a He-core white dwarf.} The vertical dashed lines mark the median value and its $1\sigma$ uncertainty.}
    \label{fig:MCMC_SED_He}
\end{figure*}

 \begin{figure*}[h]
    \centering
    \includegraphics[width=\textwidth]{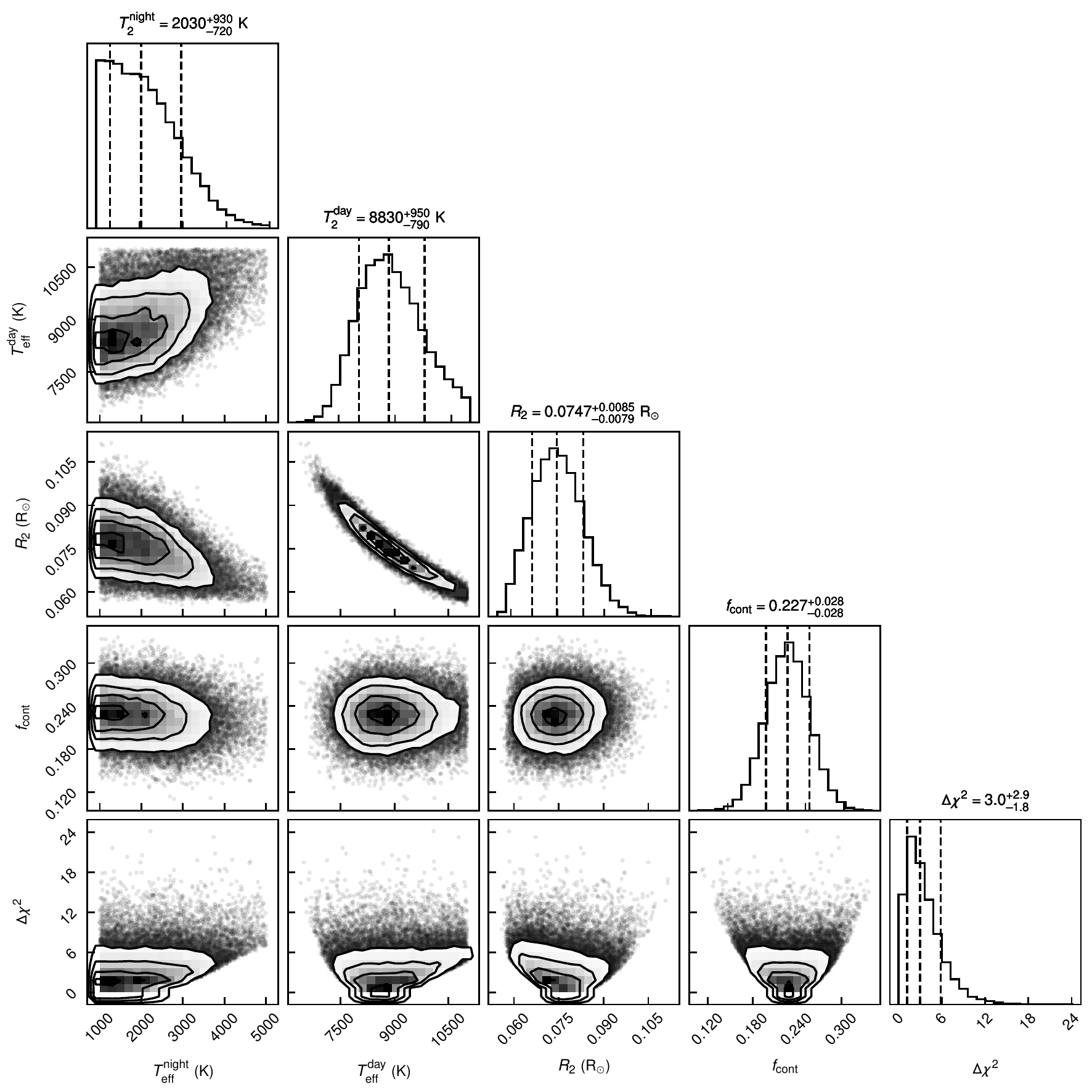}
    \caption{\textbf{One- and two-dimensional projections of the posterior probability distributions of the MCMC fit parameters for the spectral energy distribution, assuming a hybrid-core white dwarf.} The vertical dashed lines mark the median value and its $1\sigma$ uncertainty.}
    \label{fig:MCMC_SED_Hybrid}
\end{figure*}

\begin{figure}[h]
    \centering
    \includegraphics[width=\textwidth]{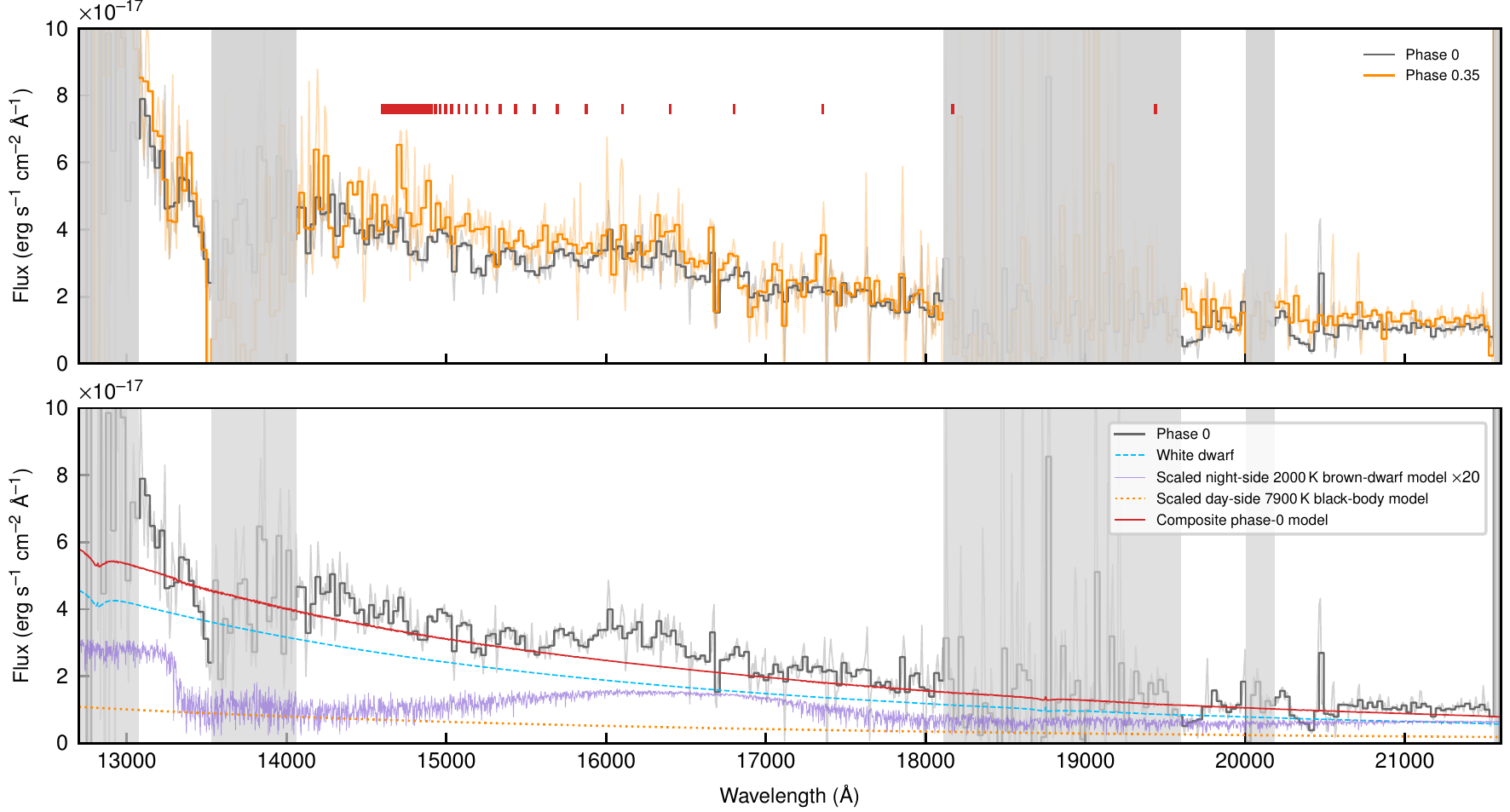}
    \caption{\textbf{Near-infrared spectra of \obj.} Binned flux-calibrated Gemini South's Flamingos-2 near-infrared spectra of \obj\ (the unbinned spectra appear as semi-transparent lines), taken near orbital phases 0 (grey) and 0.35 (orange). The red ticks mark the hydrogen Brackett series in the reference frame of the companion. The Brackett $10 \rightarrow 4$ line is possibly seen in emission at $\approx 17,357$\,\AA\ in the phase~0.35 spectrum. The greyed-out regions mark bands of high telluric atmospheric absorption. The bottom panel shows the phase-0 spectrum along with the theoretical models from Fig.~\ref{fig:SED_He}, scaled to reflect their contribution at orbital phase 0: the white-dwarf model is plotted in dashed light blue, the brown-dwarf model is plotted in solid purple (multiplied by a factor of 20 for visual clarity), the black-body model is plotted in dotted orange, and the composite model is plotted in solid red.}
    \label{fig:F2}
\end{figure}

\end{appendices}


\bibliography{WD0032}

\end{document}